\documentclass[12pt]{article}
\usepackage{graphicx}
\begin{document}
% Zeilenabstand (doppelt=8.2mm , ca. einfach:5.5mm)
\baselineskip=5.5mm
\bibliographystyle{aip}
\newcommand{\be} {\begin{equation}}
\newcommand{\ee} {\end{equation}}
\newcommand{\Be} {\begin{eqnarray}}
\newcommand{\Ee} {\end{eqnarray}}
\renewcommand{\thefootnote}{\fnsymbol{footnote}}
\def\a{\alpha}
\def\b{\beta}
\def\g{\gamma}
\def\G{\Gamma}
\def\d{\delta}
\def\D{\Delta}
\def\e{\epsilon}
\def\k{\kappa}
\def\l{\lambda}
\def\L{\Lambda}
\def\t{\tau}
\def\om{\omega}
\def\Om{\Omega}
\def\s{\sigma}
\def\lg{\langle}
\def\rg{\rangle}
\def\koff{k_{\rm off}}
\def\kon{k_{\rm on}}

\noindent
\begin{center}
{\Large {\bf Dynamic force spectroscopy: analysis of reversible bond-breaking dynamics} }\\
\vspace{0.5cm}
\noindent
{\bf Gregor Diezemann and Andreas Janshoff} \\
{\it
Institut f\"ur Physikalische Chemie, Universit\"at Mainz,
Welderweg 11, 55099 Mainz, FRG
\\}
\end{center}
\vspace{1cm}
\noindent
{\it
The problem of diffusive bond-dissociation in a double well potential under application of an external force is scrutinized.
We compute the probability distribution of rupture forces and present a detailed discussion of the influence of finite rebinding probabilities on the dynamic force spectrum. 
In particular, we focus on barrier crossing upon extension, i.e. under linearly increased load, and upon relaxation starting from completely separated bonds.
For large loading rates the rupture force and the rejoining force depend on the loading rate in the expected manner determined by the shape of the potential.
For small loading rates the mean forces obtained from pull and relax modes approach each other as the system reaches equilibrium.
We investigate the dependence of the rupture force distributions and mean rupture forces on external parameters like cantilever stiffness and influence of a soft linker. 
We find that depending on the implementation of a soft linker the equilibrium rupture force is either unaffected by the presence of the linker or changes in a predictable way with the linker-compliance. 
Additionally, we show that it is possible to extract the equilibrium constant of the on- and off-rates from the determination of the equilibrium rupture forces.
}

\vspace{0.5cm}
\vspace{1cm}
\section*{I. Introduction}
The determination of rupture forces, e.g. via atomic force microscopy (AFM) or optical tweezers by now is a standard method of investigating the strength of molecular bonds, for reviews 
see\cite{Evans:1997p969, Evans:1998p968, Evans:2001p105, Merkel:2001p2210}.  
In all the methods at hand one studies the breaking of a bond under the influence of an external force, mostly by applying a linear force ramp. 
The observed rupture forces are then monitored on a single molecule level.
Therefore, one observes a distribution of rupture forces subject to further analysis.
In most examples considered experimentally and theoretically bond rupture is treated as an irreversible event.
This means that after a rupture event the molecular configuration or structure has changed irreversibly into a very different arrangement void of memory of the prior structure. 
The analysis of either the mean rupture forces or the distribution of rupture forces yields valuable information about the details of the bond under study.
Most theoretical models that have been developed to treat this dissociation process consider it as being a problem of diffusive barrier crossing\cite{Schulten:1981p161}.
Many studies have dealt with the detailed dependence of the transition rates, i.e. the off-rates, on the applied force\cite{Bell:1978p2211, Dudko:2003p829, Hummer:2003p934, Dudko:2006p899}, on the elasticity of soft linkers\cite{Evans:1999p1079, Friedsam:2003p2214, Ray:2007p2229} and on intrinsic bond heterogeneities\cite{Raible:2006p195}.

Although most biomolecules unfold irreversibly under the influence of mechanical force there are some important exceptions.
Among these is the unfolding of RNA hairpins\cite{Manosas:2006p1126, Liphardt:2001p2264} and of ubiquitin\cite{Chyan:2004p2274} and the unzipping of coiled 
coil peptides\cite{Bornschlogl:2006p2140}. 
The reversible binding/unbinding of adhesion molecules has been studied only in recent years\cite{Seifert:2002p163, Li:2006p164}. 
Seifert, using Kramers theory, has shown that rebinding has a strong impact on the rupture force versus loading rate relation and that equilibrium is asymptotically reached for small loading rates\cite{Seifert:2002p163}. 
Only in the limit of large loading rates rebinding effects become negligible and the 
behavior expected for purely irreversible rupture is recovered.
Li and Leckband\cite{Li:2006p164} expanded the analysis by a detailed discussion of the dependence of the calculated rupture forces on kinetic parameters for one specific model potential of a double well type. 
Both approaches treat the case of a large number of equivalent adhesion molecules. 
Therefore, the force-versus-extension (FE) curves or the so-called dynamic strength 
corresponds to 'bulk' properties and the difference to properties observed in single molecule experiments has been mentioned already by Seifert\cite{Seifert:2002p163}.
A finite number of reversibly breaking adhesion bonds has been treated in a one-step master equation approach by Erdmann et al.\cite{Erdmann:2008p1530}. 
These authors presented a detailed discussion of the corresponding rupture force distribution. 
The analysis has some similarity with the one for the dissociation of sequences of identical proteins\cite{Hummer:2003p934} with the difference of finite rebinding rates. 

In the present paper we consider the reversible rupture of single molecules that exists in two states, e.g. folded or unfolded. 
We treat the problem of diffusive barrier crossing in a double well potential similar to earlier studies. 
We compare the results of our semi-analytical calculations based on the adiabatic approximation to Brownian dynamics simulations. 
In contrast to existing studies we primarily focus on the distribution of rupture forces as obtained e.g. in AFM experiments and explicitly consider the behavior of these distributions in the situation where the external force is linearly reduced. 
We discuss the dependence of the resulting distributions and rupture/rejoining forces on the model parameters and also consider the influence of flexible linkers and bond heterogeneities.
The remainder of this paper is organized as follows.
In the next section we present the theoretical calculations and section III. contains the discussion of the results. The paper closes with some concluding remarks putting emphasis in how data analysis should be carried out to avoid possible misinterpretations.
%
%\newpage
\section*{II. Theory}
We consider reversible bond breaking as a process of diffusive barrier crossing.
In the 'standard' model one considers Brownian motion in a potential of the
general form:
\be\label{V.q}
V(q,t)=V_0(q)+V_{\rm pull}(q,t)\quad\mbox{with}\quad
V_{\rm pull}(q,t)={1\over 2}k_c(q-vt)^2
\ee
Here, $q$ denotes the reaction coordinate, $k_c$ is the force constant of the 
cantilever and $v$ is the drift velocity. 
In the simplest case the force acting on the system is given by $f=k_c\cdot v\cdot t$, typically in the pN-range in AFM experiments. 
The potential $V_0(q)$ is chosen to be of a double well form with one minimum located at $q_A^0$, a transition state at $q_T^0$ and another minimum at $q_B^0$
($q_B^0>q_T^0>q_A^0$), cf. Fig.1a.
In order to study the Brownian dynamics in such a potential, one has to solve the overdamped Langevin equation associated with the potential (\ref{V.q}), given by:
\be\label{Langevin.eq}
{\dot q}(t)=-\g V'(q,t)+\xi(t)
\ee
where $\g$ is the damping constant and $\xi(t)$ is delta-correlated
Gaussian white noise,
\[
\lg\xi(t)\rg=0 \quad\mbox{and}\quad \lg\xi(t)\xi(t')\rg=2\g T\d(t-t')
\]
This Gaussian property of the noise is the reason for the fact that the stochastic
process $q(t)$ is a Markov process.
Equivalently, the process can be described by the Fokker-Planck equation
(Smoluchowski equation)\cite{vanKampen:1981, Gardiner:1997}:
\be\label{FP.eq}
{\dot p}(q,t)=D\left[{\partial\over\partial q}e^{-\b V(q,t)}
{\partial\over\partial q}e^{\b V(q,t)}\right]p(q,t)
\ee
Here, $\b=1/T$ and the diffusion constant is related to the damping constant via $D=\g T$ 
(Boltzmanns constant is set to unity throughout).
The solution of either equation has to be performed numerically, as no analytical
solutions are available for the nonlinear potentials considered here.
In this paper we will present results of Brownian dynamics simulations, i.e. the numerical solution of the Langevin equation, eq.(\ref{Langevin.eq})\cite{Branka:1998p2383}.

From the Fokker-Planck equation one can derive expressions for the transition rates
from $A$ to $T$ and from $T$ to $B$ as mean first passage times 
(MFPTs)\cite{vanKampen:1981, Gardiner:1997}:
\be\label{tau.xy.gen}
\t_{AT}(f)={1\over D}\int_{q_A}^{q_T}\!dqe^{\b V(q)}\int_{-\infty}^q\!dq'e^{-\b V(q')}
\quad\mbox{;}\quad
\t_{BT}(f)={1\over D}\int_{q_T}^{q_B}\!dqe^{\b V(q)}\int_{q}^\infty\!dq'e^{-\b V(q')}
\ee
where all quantities $q_A$, $q_T$, $q_B$ and $V(q)$ are functions of the force $f$. 
If the potential barrier is not too low, it is a good approximation for $\t_{AT}(f)$ to replace the upper limit of the inner integral by the position of the transition state, $q_T(f)$. For $\t_{BT}(f)$, one replaces the lower limit by $q_T(f)$.
This way one obtains the product approximation:
\Be\label{tau.xy.app}
\t_{AT}^{\rm product}(f)\simeq&&\hspace{-0.6cm}
D^{-1}\int_{q_A}^{q_T}\!dqe^{\b V(q)}
                             \int_{-\infty}^{q_T}\!dq'e^{-\b V(q')}
\nonumber\\
\t_{BT}^{\rm product}(f)\simeq&&\hspace{-0.6cm}
D^{-1}\int_{q_T}^{q_B}\!dqe^{\b V(q)}
                             \int_{q_T}^\infty\!dq'e^{-\b V(q')}
\Ee
The transition rates for passing from potential well $A$ to well $B$ can be calculated from the MFPTs given above using a three state model $A\leftrightarrow T\leftrightarrow B$ and assuming that the concentration of $T$ is stationary. 
The effective transition rates are given by\cite{Schulten:1981p161}:
\Be\label{k.off.on}
k^{-1}_{\rm off}(f)=&&\hspace{-0.6cm}
\t_{AT}(f)+\t_{BT}(f)Z_A(f)/Z_B(f)
\nonumber\\
k^{-1}_{\rm on}(f)=&&\hspace{-0.6cm}
\t_{BT}(f)+\t_{AT}(f)Z_B(f)/Z_A(f)
\Ee
One particular example of a double well potential that allows the analytical calculation of the MFPTs according to eq.(\ref{tau.xy.app}) is given by a harmonic cusp-like potential, cf. Fig.1a:
\be\label{V0.cusp}
V_0(q)=\left\{
	       \begin{array}{rr}
		   V_A+{1\over 2}k_A(q-q_A^0)^2 & \mbox{for $q\leq q_T^0$} \\
		   V_B+{1\over 2}k_B(q-q_B^0)^2 & \mbox{for $q\geq q_T^0$}
		   \end{array} 
		  \right.
\ee
A similar potential has been considered by Li and Leckband in their discussion of the effect of rebinding on the average dynamical strengths of molecularly bonded 
surfaces\cite{Li:2006p164}.
In eq.(\ref{V0.cusp}), the force constants are denoted by $k_A$ and $k_B$ for the left ($A$) and the right ($B$) well, respectively.
The $q_X^0$, $X=A$, $T$, $B$ are the positions of the extrema in the absence of force.
Finally, the barrier is determined by the intersection of the two parabolas,
$V_T=1/2k_A(q_T^0)^2$.
Note that in Fig.1a the overall potential according to eq.(\ref{V.q}) is plotted.
The explicit expressions for the MFPTs for the cusp-like potential given in eq.(\ref{V0.cusp}) are provided in Appendix A.
There also the Kramers rates for a general double well potential are given along with the limiting case, the rates in the Bell model\cite{Bell:1978p2211}.

In Fig.1b we show the rates $\koff(f)$ and $\kon(f)$ resulting from the exact expression, eq.(\ref{tau.xy.gen}), from the product approximation (\ref{tau.xy.app}) and the Kramers approximation, eq.(\ref{tau.xy.kramers}).
It is evident that the product approximation yields excellent results, whereas the Kramers approximation fails when the force at which well A vanishes is approached.

Denoting the time-dependent population of well $A$ by $n(t)$, then the master equation reads after a transformation to the force:
\be\label{ME.f}
{d\over df}n(f)=-\left[\k_{\rm off}(f)+\k_{\rm on}(f)\right]n(f)+\k_{\rm on}(f)
\ee
Here, we defined $\k_{\rm on/off}(f)=(df/dt)^{-1}k_{\rm on/off}(f)$. 
If one assumes that the time-dependence of the force is determined solely by the pulling, 
$f=\mu_c\cdot t$, the loading rate $(df/dt)=\mu_c$ reads $\mu_c=k_c\cdot v$.
However, the situation becomes more involved if a soft linker such as a polymer, long nucleic acid, polysaccharide or polypeptide chain is coupled to the bond along the reaction coordinate\cite{Hummer:2003p934,Evans:1999p1079}.
In principle, the energy landscape can no longer be considered as one-dimensional, as the linker has to be treated explicitly.
An example of a calculation for the escape from a single well has been presented in ref.\cite{Sain:2004p2382}.
In the present paper, we will discuss two approximative ways of treating the impact of a soft linker on the force spectrum.
In the model of Evans and Ritchie\cite{Evans:1999p1079} the change in the compliance of the system due to the presence of a soft linker gives rise to a modified loading rate:
\be\label{mu.WLC}
(df/dt)^{-1}=\mu_c^{-1}\left[1+C_{L}(f)\right]
\ee
with the bare cantilever loading rate $\mu_c=k_c\cdot v$ and the force-dependent compliance of the linker $C_L$.
We will restrict ourselves to the worm-like chain (WLC) model for the description of a soft linker.
In this case one has $C_{WLC}(f)=2\b l_pL_ck_c(1-z(f))^3/[1+2(1-z(f))^3]$. In this expression, $z(f)$ is the force-dependent extension of the WLC-linker, i.e. $z(f)$ is determined by the force law $f(z)=(1/4\b l_p)z/(1-z)^2(6-9z+4z^2)$. 
Furthermore, $l_p$ and $L_c$ denote the persistence length and the contour length of the polymeric linker, respectively.

Another way to treat the effect of a soft linker is the introduction of a potential of mean force where the linker coordinate is integrated out\cite{Hummer:2003p934}.
In this approach one considers the potential consisting of the sum of the potential given in eq.(\ref{V.q}), $V_0(q)+V_{\rm pull}(q,t)$, and the linker potential $V_{\rm link}(q-q_l)$
with the linker coordinate $q_l$. The potential of mean force is defined by the relation
$e^{-\b V_{\rm eff}(q,t)}
=\int\!dq_le^{-\b(V_0(q)+V_{\rm pull}(q,t)+V_{\rm link}(q-q_l))}$.
In a harmonic approximation, this procedure results in a potential of the same form as given in eq.(\ref{V.q}) but with an effective force constant for the cantilever-linker system,
\be\label{kc.eff}
k_{\rm eff}=(k_c^{-1}+k_L^{-1})^{-1}
\ee
Here, $k_L$ denotes the linker force constant. 
Therefore, in this approach one has to replace the bare cantilever force constant $k_c$ by a smaller $k_{\rm eff}$. 
We mention that for the WLC model the force constant of the linker is given by 
$k_{WLC}=3/(2\b l_pL_c)$.

Note that in eq.(\ref{ME.f}) we already used the fact that the sum of the populations in well $A$ and well $B$ equals unity.
The solution of the master equation is standard and can be performed in a semi-analytical manner. Knowledge of $n(f)$ allows the computation of all quantities of interest. 
(If one is interested in the pull mode, one uses the natural initial condition $n(f_0)=1$ and in case of the relax mode, starting from a (high) force $f_{\rm max}$, this is replaced by $n(f_{\rm max})=0$.)

In the standard description of dynamic force spectroscopy, one considers the rupture force distribution which is given by $p(f)=-n'(f)$ with the prime denoting the derivative with respect to the force $f$.
Note that this definition usually is employed in the absence of rebinding. 
However, we will use the same definition in case of finite $k_{\rm on}$, cf. ref.\cite{Evans:2001p105}.
Hence, the rupture force distribution is given by:
\be\label{pf.def}
p(f)=-{d\over df}n(f)=\left[\k_{\rm off}(f)+\k_{\rm on}(f)\right]n(f)-\k_{\rm on}(f)
\ee
Using this distribution one can calculate the moments
\be\label{moments.gen}
\lg f_r^n\rg=\int_0^\infty\!dff^np(f)=n\int_0^\infty\!dff^{n-1}n(f)
\ee
The most important ones, the mean rupture force $\lg f\rg$ and the second moment 
$\s_r^2=\lg(f_r-\lg f_r\rg)^2\rg$ will be discussed later. 
We will show that the mean rupture force defined by eq.(\ref{moments.gen}),
$\lg f_r\rg$, behaves very similar to the maximum of $p(f)$, $f_{r,m}$.

As mentioned already in the Introduction, Seifert\cite{Seifert:2002p163} and also Li and Leckband\cite{Li:2006p164} considered a so-called dynamic strength, given by:
\be\label{DS.def}
\hat F(f)=k_c\lg\left[v\cdot t-q\right]\rg_A
\ee
where $\lg\left[v\cdot t-q\right]\rg_A
=Z_A^{-1}\int_A\!dq\left[v\cdot t-q\right]e^{-\b V(q,f)}\cdot n(f)$ and analogously for well $B$.
Here, we will consider the averaged FE-curve
\be\label{FvsX.def}
F(f)=k_c\left[\lg\left[v\cdot t-q\right]\rg_A
           +\lg\left[v\cdot t-q\right]\rg_B\right]
\ee
because this can be obtained directly from simulations (and actual measurements near equilibrium\cite{Bornschlogl:2006p2140}).
In a Gaussian approximation for the averages, cf. eq.(\ref{qf.gauss}), one finds:
\be\label{FvsX.res}
F(f)\simeq[f-k_c q_A(f)]n(f)+[f-k_c q_B(f)](1-n(f))
\ee
where $q_X(f)=(k_Xq_X^0+f)/(k_X+k_c)$ is the equilibrium position of well $X$ under the influence of the external force.
We will also briefly discuss the substantial differences between the local maximum of the averaged FE-curve, $F_m$, and the maximum of the rupture force distribution, $f_{r,m}$.
\section*{III. Results and discussion}
\subsubsection*{A. Rupture force distributions and dynamic strength}
We start the discussion with the treatment of the Brownian motion in a potential of the form given in eq.(\ref{V0.cusp}).
In Fig.2a we show typical examples of FE-curves as obtained from Brownian dynamics simulations which were carried out employing a stochastic Runge-Kutta 
algorithm\cite{Branka:1998p2383}.
All parameters used in the simulation are given in the caption to Fig.1a. 
These curves can be viewed as representative for experimentally observed FE-curves in the case of a hard linker.
We have chosen a set of parameters for the potential closely related to what is known for the strength of a single hydrogen bond (Fig.1a). 
Reversibility on experimentally relevant time scales was realized by placing well $B$ energetically close to the transition state rendering the system irreversible only at high loading rates.

It is evident that the rupture force and the rejoining force are stochastic quantities. 
Of course, whenever a rupture or rejoining event takes place, the system crosses the position of the transition state. This position is independent of force in our example, $q_T(f)=q_T^0$, because we consider a steep barrier. 
This does not hold if a soft barrier is considered instead. 
We will later briefly discuss the case of a finite curvature of the potential at $q_T^0$. 
All further information about the system, such as rupture force distributions or averaged FE-curves, is based on an analysis of simulated FE-curves.

In order to check the correctness of our definition of the rupture force distribution, eq.(\ref{pf.def}), we compare the calculated distributions $p(f)$ with the one obtained directly from the simulations (Fig.2b). 
It is evident that both methods yield the same results for highloading rates. 
At smaller loading rates the situation becomes more complicated due to fluctuations in the force FE-curves rendering the correct determination of the rupture event difficult. 
A better way to validate the theory is hence to compare averaged FE-curves as demonstrated in Fig.2c.
Perfect agreement between simulations and theory was found, as previously reported by Li and Leckband albeit we always use the factorization approximation for the MFPTs, eq.(\ref{tau.xy.app}).
It also can be seen immediately that the mean rupture force increases as a function of loading rate in the pull mode and decreases in the relax mode, as has been found 
earlier\cite{Manosas:2006p1126}.
Averaged FE-curves have been obtained by averaging 1000 different individual FE-curves obtained by simulations, cf. Fig.2a. 
The dashed lines in Fig.2c are the Gaussian approximation to the average FE-curves, eq.(\ref{FvsX.res}). 
We found that the agreement between the Gaussian approximation and the simulated curves decreases somewhat with increasing loading rate. However, for a loading rate as high as $\mu_c=3\cdot 10^5$ pN/s the maximum discrepancy is on the order of 5\%.
Given this satisfying agreement between simulation results and semi-analytical calculations, in the following we will solely discuss the results of the solution of the master equation (\ref{ME.f}), i.e. the distributions $p(f)$.

In Fig.3a, we show rupture force distributions for various loading rates, both in the pull mode (upper panel) and the relax mode (lower panel). 
It is evident that the skewness of the distributions in the relax mode is just opposite to the one in the pull mode. 
Furthermore, we included the distributions for vanishing rebinding (dash-dotted lines). For loading rates $\mu_c>10^4$ pN/s rebinding effects can safely be neglected for the parameters chosen.
For small loading rates, however, rebinding is essential as it prevents the maximum of 
$p(f)$ to reach zero instead of its equilibrium value.

The rupture forces and rejoining forces are plotted versus loading rate in the upper panel of Fig.3b. It is evident how the values from the pull and the relax mode converge to the equilibrium value for small loading rates.
It is also obvious that the overall behavior of $\lg f_r\rg$ and $f_{r,m}$ is roughly the same.
This does not hold for the mean rupture forces in the absence of rebinding, $\lg f_r\rg_0$ (dotted lines).
For small loading rates these behave qualitatively different, as equilibrium is absent. 
Only for large loading rates the effects of rebinding becomes negligible and 
$\lg f_r\rg\simeq\lg f_r\rg_0$. 
We mention that we calculated $\lg f_r\rg_0$ in the pull mode by simply setting $\kon=0$ in eq.(\ref{ME.f}). 
In Appendix C, it is briefly explained how the present off-rate is related to the one used
in the calculation performed by Dudko et al.\cite{Dudko:2006p899}.

In the lower panel of Fig.3b, we show the width of the rupture force distribution, 
$\s_r=(\lg f_r^2\rg-\lg f_r\rg^2)^{1/2}$.
The fact that $\s_r$ is nearly constant for small $\mu_c$ clearly reflects the influence of the finite rebinding probability.
Only in the regime where rebinding becomes less important the width increases and for large 
$\mu_c$ coincides with the corresponding one for vanishing rebinding (dotted lines).
Therefore, in the interesting $\mu_c$-regime, the width looks similar to what is predicted by the Bell-model without rebinding\cite{Dudko:2006p899}. 
Consequently, $\s_r$ data should be treated with care.
(The fact that the apparent slope changes for small $\mu_c$ in the pull mode and for large 
$\mu_c$ in relax mode is due to the fact that we only considered positive forces in our calculations similar to typical experimental situations.)

As noted above, the rupture force distribution $p(f)$ and the dynamic strength 
$\hat F(f)$ give rise to different maximum values for the 'rupture' force or the 'characteristic' force. 
This finding is a manifestation of the observation already made by Seifert\cite{Seifert:2002p163} that the two forces $f_{r,m}$ and $\hat F_m$ are distinct quantities. 
We note that the differences between the averaged FE-curves and the dynamic strength depend on the loading rate. In Fig.3c we plot the Gaussian approximation for $F(f)$ versus the extension $q$ for several loading rate and in the inset this is compared to $\hat F(f)$. 
In order to quantify the difference, we calculate various characteristic force values which are plotted in Fig.3d versus loading rate $\mu_c$. 
As indicated in Fig.3c, in the pull mode the $\hat F_m$-values are smaller than $F_m$ and these in turn are smaller than $f_{r,m}$. 
When the data are shifted to start at the same low-$\mu_c$ limit, one can see the different apparent slopes of the various characteristic force values. 
For large $\mu_c$, the difference in slope between $f_{r,m}$ and $\hat F_m$ is on the order of 25\%.
In the relax mode, the differences between the various values are much less outspoken, mainly due to their smaller overall variation. 
From an experimental point of view, systems close to equilibrium displaying FE-curves fluctuating between state $A$ and $B$ are better analyzed in terms of the averaged FE-curves instead of the rupture force distributions $p(f)$ by recording merely the last rupture or rejoining event, as already mentioned by Evans and Ritchie\cite{Evans:1997p969}. 
Of course, the differences between the characteristic forces $F_m$ and $f_{r,m}$ have to be taken into account.\\

\noindent
We close this section with noting that the differences in the 'rupture' and 'rejoining' forces determined from averaged FE-curves and from rupture and rejoining force distributions have their origin in the intrinsic nonlinearity of all force-dependent quantities. This also holds in equilibrium, see Fig.3d. 
We mention that it is not simply related to the non-gaussian shape of $p(f)$, as can be shown by calculating $F(f)$ with an assumed Gaussian $n(f)$ (not shown). 

In the following we discuss the dependence of experimentally relevant properties such as rupture and rejoining forces on kinetic parameters. We mainly focus on such quantities that have not already been discussed earlier by Li and Leckband for the dynamic strength\cite{Li:2006p164}. 
However, in the following, we will use Kramers rates for a general bistable potential in order to be able to also discuss the influence of a soft barrier. 
We will focus on the distribution of rupture forces and quantities derived from $p(f)$.
\subsubsection*{B: Variation of kinetic parameters}
In order to be able to include a discussion of the effect of a soft barrier and not to be restricted to the quite special form of the potential discussed above, in this section we consider a more general form of a double-well potential and apply Kramers theory to the calculation of the transition rates. 
Without going into details, we give the results for the off- and on-rates in Appendix A, 
eq.(\ref{koff.Kramers}).
The parameters can be chosen in such a way that both, the shape of the rupture force distributions and the mean values coincide with those obtained for the cusp-like potential discussed above. 
This is demonstrated in Fig.4a, where $\lg f_r\rg$ is plotted versus $\mu_c$. 
Also included in that figure is the result of a calculation with rates according to the Bell model, cf. eq.(\ref{koff.Bell})\cite{Bell:1978p2211}. 
The discrepancy between the latter and the local harmonic or the cusp model is similar to what is observed in case of a pure escape model\cite{Dudko:2006p899}. 

Next, we consider the parameters relevant for the characterization of the transition state in the local harmonic approximation, namely the position $q_T^0$ and the curvature $k_T$. 
The mean rupture forces $\lg f_r\rg$ are plotted versus $\mu_c$ in Fig.4b for different values of these parameters. 
In the upper panel $q_T^0$ is varied and it is obvious that the effect is just what one expects from the Bell model, namely the $\lg f_r\rg(\rm log(\mu_c))$-curves become steeper for smaller $q_T^0$. 
We note that the discrepancy between the local harmonic approximation and the Bell model
strongly depends on the distance of the well to the transition state. In order to show this, we included results from calculations using the Bell expressions, eq.(\ref{koff.Bell}) as the dotted lines in Fig.4b.
It is evident that the discrepancy diminishes with increasing $q_T^0$. 
The reason for this behavior is easily understood from the derivation of the Bell expressions for the transition rates as the limit of large critical force $f_c$ of the Kramers rate, cf. 
eq.(\ref{koff.Bell}), because $f_c$ is directly proportional to $q_T^0$.
Explicitly, one has for large $\mu_c$, where rebinding is negligible, cf. e.g.\cite{Dudko:2006p899}:
\be\label{fr.pull.relax.Bell}
\lg f_r\rg_{\rm pull}\simeq {\ln{(z_p\mu_c)}\over Q_{TA}}
\quad\mbox{and}\quad
\lg f_r\rg_{\rm relax}\simeq {-\ln{(z_r\mu_c)}\over Q_{BT}}
\ee
with $z_p=e^{-\g}Q_{TA}/\koff^0$, $z_r=e^{-\g}Q_{BT}/\kon^0$, $\g=0.577215...$ being the Euler constant. 
Of course, in the relax mode the effect is very small for the parameters chosen.

This situation changes when a variation in curvature is considered (Fig.4b, lower panel). 
In the Bell model, the on- and off-rates and therefore also the rupture and rejoining forces are independent of the value of $k_T$. 
Here, we find an increasing apparent slope at large loading rates for decreasing $k_T$ because we are using the on- and off-rates in the Kramers approximation, eq.(\ref{koff.Kramers}).
At small loading rates, however, $\lg f_r\rg$ is hardly affected at all. 
In the relax mode, no changes in the rejoining force are found for the present parameters. 
This behavior can be understood qualitatively from the expressions for the scales of the relevant forces given in eq.(\ref{fc.fr}). 
For the off-rate this scale is given by $f_c\simeq {k_Tk_A\over k_A+k_T}q_T^0$ and thus directly proportional to $k_T$.
In the relax mode, the relevant rate is the on-rate and here one has for the scale
$f_r\simeq (k_B+k_c)(q_B^0-q_T^0)$ independent of $k_T$. 
(Note that this argument holds for soft $B$-wells only.)

The dependences of the characteristic scales $f_c$ and $f_r$ on the various force constants also allows to qualitatively understand the dependence of the rupture and rejoining forces on the cantilever spring constant $k_c$ for large $\mu_c$.
Again assuming $(k_A\sim k_T)\gg (k_B\sim k_c)$ we have almost no dependence of $\koff(f)$ on $k_c$ whereas $\kon(f)$ changes strongly with $k_c$. 
Thus, for large $\mu_c$ we expect a very weak $k_c$ dependence of $\lg f_r\rg$ in the pull mode and a much stronger one in the relax mode. 
This is just what is observed in the calculation as shown in Fig.4c. In the upper panel one can see that for large $\mu_c$ the behavior just discussed is found, whereas for small $\mu_c$ there is strong dependence on $k_c$, also in the pull mode. 
In the lower panel, the dependence of the mean rupture force in equilibrium ($\mu_c=0$) on $k_c$ is displayed.
For comparison, also the results for the Bell model are shown as the dotted lines.
In case of this model the rates are given by eq.(\ref{koff.Bell}) and the mean rupture force can be calculated analytically (cf. eq.(\ref{moments.gen})):
\be\label{fr.eq.Bell}
\lg f_r\rg^{\rm eq}=\int_0^\infty\!dfn^{\rm eq}(f)=\int_0^\infty\!{df\over 1+K_{\rm eq}^{-1}(f)}
={1\over Q_{BA}}\ln{\left(1+K_{\rm eq}(0)\right)}
\ee
where $K_{\rm eq}(f)=\kon(f)/\koff(f)$. 
Similarly, one finds $f_{r,m}^{\rm eq}=\ln{\left(K_{\rm eq}(0)\right)}/Q_{BA}$.
It is important to note that $K_{\rm eq}(0)$ for the parameters chosen depends on $k_c$.
The same holds for $Q_{BA}=\b k_B/(k_B+k_c)q_B^0$. The upper dotted line has been calculated using this expression and the lower one with the approximation $Q_{BA}=\b q_B^0$. 
(If one uses the $k_c=0$-value for $K_{\rm eq}(0)$ along with $Q_{BA}/\b=1$ nm, one finds 
$\lg f_r\rg^{\rm eq}\simeq 26.3$ pN for the parameters chosen.)
One can readily see that the distance between the two minima can be extracted from 
$\lg f_r\rg^{\rm eq}$ data if these are extrapolated to $k_c\to 0$. In this case all three curves yield a distance between $q_A^0$ and $q_B^0$ of about $1$ nm within 20\% accuracy. 
Thus, if one has an estimate of the equilibrium constant $K_{\rm eq}(0)$ from other kinetic measurements, it is possible to get information about the position of the minima $q_A^0$ and $q_B^0$ from the equilibrium mean rupture forces. 
On the other hand, as the mean rupture forces in the large $\mu_c$-limit allow estimates of 
$Q_{TA}$ and $Q_{BT}$, cf. eq.(\ref{fr.pull.relax.Bell}), one can use the $k_c$-dependence of the equilibrium mean rupture force in order to determine $K_{\rm eq}$.
\subsubsection*{C: Impact of a soft linker}
As has been mentioned in the preceeding section, one frequently is confronted with a situation where the influence of a soft linker cannot be neglected. 
In the model considered by Evans and Ritchie\cite{Evans:1999p1079}, for vanishing rebinding the main effect of a soft linker is to broaden the rupture force distributions somewhat and to decrease the mean rupture forces due to the smaller effective force constant of the cantilever-linker system. 
We use eq.(\ref{mu.WLC}) in order to treat a soft WLC-linker in conjunction with the master equation, eq.(\ref{ME.f}), for finite rebinding rates.
In Fig.5a, we show the influence of polymeric linkers of varying contour length $L_c$ on the rupture force distributions. 
It is seen that the effect of the smaller compliance of the linker is very similar to what has been found in the absence of rebinding in the pull mode, i.e. a shift of the peak maximum towards smaller forces. 
In the relax mode just the opposite shift is found, as expected. 
Therefore, the mean rupture forces determined in the two modes shift together and the equilibrium force is reached at a somewhat higher loading rate than in case of a hard linker. 
Note the distortion of the $p(f)$ curves indicating more weight at lower (pull) or higher (relax) forces. It thus appears that a soft linker has some impact on the skewness of the rupture force distributions.

The shift in the mean rupture forces is shown in Fig.5b, where we plot $\lg f_r\rg$ versus $\mu_c$ for a hard linker ($L_c=0$) and a WLC-linker with a contour length of $50$ nm and a persistence length of $0.4$ nm. Also the fact that the rupture force in equilibrium, $\lg f_r\rg^{\rm eq}$, is independent of the linker is obvious.
This fact can easily be understood from the equilibrium solution of the master equation, cf. eq.(\ref{fr.eq.Bell}), $n(f)^{\rm eq}(f)=[1+K_{\rm eq}^{-1}(f)]^{-1}$ with
$K_{\rm eq}(f)=\k_{\rm on}(f)/\k_{\rm off}(f)=\kon(f)/\koff(f)$ independent of $(df/dt)$ and thus of $C_L(f)$. 
Therefore, the rupture force $\lg f_r\rg^{\rm eq}$ is unaffected by the influence of a soft linker  as observed for the B-S transition of DNA\cite{ClausenSchaumann:2000p2526}. 

However, we note that this is not a generally valid result. 
If we treat the influence of a soft linker in the manner discussed by Hummer and Szabo\cite{Hummer:2003p934}, we have to use a small effective force constant given by eq.(\ref{kc.eff}) instead of the bare cantilever force constant. 
The dotted lines in Fig.5b are obtained this way and one clearly sees a behavior similar to that discussed in the context of the $k_c$-dependence of $\lg f_r\rg^{\rm eq}$, cf. Fig.4c.

From this discussion we conclude that measurements of the equilibrium rupture force can in favourable cases be used in order to improve the understanding of the impact of soft linkers on the force spectrum.
\subsubsection*{D: Bond heterogeneities}
It has been shown that in some cases the experimentally observed skewness of the rupture force distributions is opposite to what is expected on the basis of model calculations, independent of the model used\cite{Raible:2006p195, Raible:2006p178}. The explanation put forward by Raible et al. is that the parameters entering the off-rate are distributed. 
In particular, if the off-rate is written as $\koff(f)=\koff^0e^{Q_{TA}f}$, a distribution of the distance to the transition state, $Q_{TA}$, is considered and this is compatible with the experimental findings for a number of examples\cite{Raible:2006p178}. 

If one considers a finite rebinding probability with a rate $\kon(f)=\kon^0e^{-Q_{BT}f}$ one has two parameters $Q_{TA}$ and $Q_{BT}$ that can be considered as distributed.
We have performed calculations of the rupture force distributions and the mean rupture forces with distributions of both parameters. 
Without showing results here, we mention that the effect is very similar to what has been observed in the case without rebinding. 
The widths of the rupture force distributions are more strongly affected by bond heterogeneities in the pull mode than in the relax mode. 
The mean rupture forces are hardly changed at all.
\section*{III. Conclusions}
The model calculations presented in the present paper demonstrate the importance of finite rebinding probabilities in the interpretation of dynamic force spectroscopy experiments. 
We have shown that the usual definition of the rupture force distribution also is meaningful if rebinding cannot be neglected. 
Therefore, one can discuss the so-called force spectrum in the same way as it is usually done when one is concerned with irreversible bond breaking events. 
There is, however, a difference to the dynamic strength, a bulk property, that has been considered in earlier work on reversible kinetics in the context of dynamic force spectroscopy. 
From an experimental point of view, however, near equilibrium fluctuations in the individual FE-curves can prevent correct data aquisition for rupture and rejoining force histograms.
In such cases it may be more practicable to record averaged FE-curves as shown by 
Bornschl\"ogl and Rief\cite{Bornschlogl:2006p2140}.

One effect of finite on-rates regards the interpretation of the width of the rupture force distributions. It has been shown earlier\cite{Dudko:2006p899} that this width increases with increasing loading rate without rebinding. 
Here, we find that due to finite on-rates the width is almost constant over a larger range of loading rates and only for large $\mu_c$ starts to increase in the same way as without rebinding. 
Therefore, in order to discriminate among various models for the transition rates one has to carefully cover a large range of loading rates.

We expect that it is not easy to discriminate between the Bell model and more complex scenarios for distances to the transition state typical for proteins. 
This is due to the fact that the Bell-like rates are obtained from those for a local harmonic approximation in the limit of large critical forces. 
These critical forces themselves are proportional to the distance to the transition state.
For larger distance the force spectrum shows less curvature, cf. Fig.4b.  

It has been shown that the determination of the equilibrium rupture force allows the extraction of the equilibrium constant $K_{\rm eq}$ at zero load. This quantitiy, however, depends on the force constant of the cantilever and should therefore be extrapolated to $k_c\to 0$ in order to be able to compare it directly to other determinations like bulk measurements. 
Small cantilever force constants are generally recommended alos to minimize the impact of polymeric linkers and their variation in contour length on the outcome of a force experiment.

We investigated the impact of a soft linker on the rupture force distributions and the mean rupture force.
We found that in principle it should be possible from detailed determinations of the equilibrium rupture force to gain some deeper understanding of the compliant behavior of the composite cantilever-linker system. 

In conclusion, we have shown that dynamic force spectroscopy can yield insightful results on the nature of system showing reversible bond-breaking and that the equilibrium regime of small loading rates is particularly interesting in this respect.
%greg
%
%\newpage
%
\begin{appendix}
\section*{Appendix A: Transition rates}
\subsubsection*{MFPTs for the harmonic cusp-like potential}
\setcounter{equation}{0}
\renewcommand{\theequation}{A.\arabic{equation}}
In this Appendix, the explicit expressions for the transition rates for the cusp-like potential and the Kramers rates in a general potential are given.
For a potential built from two harmonic wells with a sharp intersection on finds from eq.(\ref{tau.xy.app}):
\be\label{tau.ZX.IXT}
\t_{AT}^{\rm product}(f)=D^{-1}Z_AI_{AT}
\quad\mbox{and}\quad
\t_{BT}^{\rm product}(f)=D^{-1}Z_BI_{BT}
\ee
with
\be\label{Z.X}
Z_X(f)=e^{-\b V_X(f)}\left[\left({2\pi\over \b k_{XC}}\right)^{1/2}
-\left({\pi\over 2\b k_{XC}}\right)^{1/2}{\rm erfc}\left\{\D_{XT}(f)\right\}
 \right]
\quad\mbox{; $X=A$, $B$}
\ee
Here, we used the abbreviations $k_{XC}=k_X+k_c$,
\[
V_X(f)=V_X-{1\over 2}k_{XC}q_X(f)^2+{1\over 2}k_X(q_X^0)^2
\quad\mbox{with}\quad q_X(f)=k_{XC}^{-1}(k_Xq_X^0+f)
\]
and
\[
\D_{AT}=\left({\b k_{AC}\over 2}\right)^{1/2}\left[ q_T(f)-q_A(f)\right]
\quad;\quad
\D_{BT}=\left({\b k_{BC}\over 2}\right)^{1/2}\left[ q_B(f)-q_T(f)\right]
\]
Furthermore, we defined
\be\label{I.XT}
I_{XT}(f)=\left({2\over \b k_{XC}}\right)^{1/2}e^{\b V_X(f)}
{\rm d}\left(\D_{XT}(f)\right)
\quad\mbox{with}\quad
{\rm d}(x)=\int_0^x\!dze^{z^2}\simeq{e^{x^2}\over 2x}
\ee
Using eq.(\ref{tau.ZX.IXT}) yields results for the populations of the potential wells as a function of force (time) that are indistinguishable from calculations with the exact expression for the MFPTs.

The Kramers approximation for the MFPTs is obtained from eq.(\ref{tau.ZX.IXT}) if one keeps only the first term in eq.(\ref{Z.X}) and using ${\rm d}(x)\simeq{e^{x^2}\over 2x}$:
\be\label{tau.xy.kramers}
\t_{XT}^{\rm Kramers}=\left({\sqrt{\pi}\over\b D k_{XC}}\right){\exp{(\D_{XT}^2)}\over\D_{XT}}
\ee
\subsubsection*{Transition rates in a local harmonic approximation}
Next, we give the expressions for the Kramers rates for off- and on-rates in a general double well potential\cite{vanKampen:1981, Gardiner:1997}.
Approximating the minima as well as the maxium by a parabola with curvatures $k_A$, $k_B$ for the minima and $k_T$ for the maximum, one finds:
\be\label{koff.Kramers}
\koff(f)=\koff^0e^{\b Q_{TA}f\left[1-{1\over 2}f/f_c\right]}
\quad\mbox{and}\quad
\kon(f)=\kon^0e^{-\b Q_{BT}f\left[1+{1\over 2}f/f_r\right]}
\ee
Here, $Q_{TA}=Q_T-Q_A$, $Q_{BT}=Q_B-Q_T$ with $Q_X=k_X/k_{XC}q_X^0$ for $X=A$, $B$ and $Q_T=k_T/(k_T-k_c)q_T^0$. 
Furthermore, the scale of the forces are set by
\be\label{fc.fr}
f_c={(k_T-k_c)k_{AC}\over k_A+k_T}Q_{TA}
\quad\mbox{and}\quad
f_r={(k_T-k_c)k_{BC}\over k_B+k_T}Q_{BT}
\ee
and the rates in the absence of any force are given by:
\be\label{koff.0}
\koff^0=D{\sqrt{(k_T-k_c)k_{AC}}\over 2\pi}e^{-\b(V_T-V_A)}
\quad\mbox{and}\quad
k_{\rm on}^0=D{\sqrt{(k_T-k_c)k_{BC}}\over 2\pi}e^{-\b(V_T-V_B)}
\ee
Finally, it easy to show from eq.(\ref{koff.Kramers}) that in the limit of large $f_c$ and $f_r$ the rates of the Bell model 
\be\label{koff.Bell}
\koff^{(B)}(f)=\koff^0e^{\b Q_{TA}f}
\quad\mbox{and}\quad
\kon^{(B)}(f)=\kon^0e^{-\b Q_{BT}f}
\ee
are obtained. 
Here, $Q_{TA}$ and $Q_{BT}$ in the limit $k_c\to 0$ are the respective distances to the transition state.
\section*{Appendix B: Averaged force versus extension curves}
\setcounter{equation}{0}
\renewcommand{\theequation}{B.\arabic{equation}}
Here, we derive the expressions for the averaged FE-curves from the rupture force distribution.
In a Gaussian approximation, the expectation values $\lg q(f)\rg$ are given by 
\be\label{qf.gauss}
\lg q(f)\rg_X=Z_X^{-1}\int_X\!dq e^{-\b V(q,f)}q
=q_X(f)={k_Xq_X^0+f\over k_{XC}}
\ee
with $X=A$ or $X=B$ depending on the potential well considered.
In this approximation the increase in the force as a function of the extention is simply given by $[f-k_cq_X(f)]$, which of course is the same as the force measured for instance by an AFM cantilever as long as the bond considered is intact.
With this, the overall force is given by:
\be\label{F.fr}
F=\left\{
	       \begin{array}{rr}
		   f-k_c q_A(f) & \mbox{for $f<f_r$} \\
		   f-k_c q_B(f) & \mbox{for $f>f_r$}
		   \end{array} 
		  \right.
\ee
This means, one can write
\be\label{F.ges.def}
F(f,f_r)=[f-k_c q_A(f)]\Theta(f_r-f)+[f-k_c q_B(f)]\Theta(f-f_r)
\ee
where $\Theta(x)$ denotes the Haeviside step function, $\Theta(x)=1$ if $x\geq 1$ and $\Theta=0$ if $x<0$.
Here, $f_r$ is the rupture force and eq.(\ref{F.ges.def}) is just an expression of the fact that after rupture the system is in the right hand minimum '$B$' of the potential.
The average over the rupture forces is given by the rupture force distribution, i.e.
\be\label{F.ges.mit}
\lg F(f)\rg=\int_0^\infty\!df_r F(f,f_r)p(f_r)
=[f-k_c q_A(f)]n(f)+[f-k_c q_B(f)](1-n(f))
\ee
which follows immediately from the definition of $\Theta(x)$ and the fact that 
$p(f)=-{d\over df}n(f)$.

\section*{Appendix C: Escape from a harmonic well}
\setcounter{equation}{0}
\renewcommand{\theequation}{C.\arabic{equation}}
Here, we discuss the transition from a double well potential to a model of escaping a single harmonic well as discussed by Dudko et al.\cite{Dudko:2006p899}. In the product approximation of eq.(\ref{tau.xy.app}) one has for the off-rate, cf. eq.(\ref{k.off.on}):
\be\label{k.off.tau}
k^{-1}_{\rm off}(f)=\t_{AT}(f)+\t_{BT}(f)Z_A(f)/Z_B(f)
\simeq D^{-1}Z_A\left[I_{AT}+I_{BT} \right]
\ee
If the potential is given by the sum of eq.(\ref{V0.cusp}) and the pulling potential according to eq.(\ref{V.q}), the limiting case of a single well adjacent to an unbound state is provided by the limit $V_B\to(-\infty)$.
If one chooses a fixed value for the force constant $k_B$, the condition of the crossing point of the two parabola yields
$q_B^0=q_T^0+([2(V_T-V_B)-k_c(q_T^0)^2]/k_B)^{1/2}$
where $q_T^0$ is defined by $V_T=1/2k_{AC}(q_T^0)^2$. 
For large negative $V_B$ one then has $V_B\simeq -1/2k_B(q_B^0)^2$.
If one now uses eqns.(\ref{Z.X},\ref{I.XT}) one finds that
\be\label{k.off.tau.at}
I_{BT}\simeq(\b k_Bq_B^0)^{-1}
\quad\mbox{and thus}\quad
k^{-1}_{\rm off}(f)\simeq\t_{AT}(f)
\ee
The resulting rate for escape from a harmonic well, $k_s(f)$, can be cast into the 
form\cite{Dudko:2006p899}
\be\label{k.swp.f}
k_s(f)=k_s(0)\e e^{\b V_T(1-\e^2)}
\quad\mbox{with}\quad
k_s(0)=D{(\b k_{AC})^{3/2}\over\sqrt{2\pi}}q_T^0e^{-\b V_T}
\ee
where the abbreviation $\e=1-f/(k_{AC}q_T^0)$ has been used and $V_T$ is given above.
\end{appendix}
\section*{Acknowledgment}
We thank Thorsten Metzroth, Matthias Janke, Ingo Mey and J\"urgen Gauss for fruitful discussions.
Financial support by the Deutsche Forschungsgemeinschaft via the SFB 625 is acknowledged.
\newpage
\section*{Figure captions}
\begin{description}
\item[Fig.1 ] (color online)\\
{\bf a:}
A sketch of a double well potential as used in the calculations.
the explicit parameters are chosen as follows:
The cantilever force constant was set to $k_c=30$ pN/nm, $k_A=1000.0$ pN/nm and 
$k_B=48.49$ pN/nm;
the diffusion constant is $D=2000.0$ nm$^2$/s; 
the potential heights are chosen as $V_A=0$, $V_B=8.0/\b=33.12$ pNnm and from this one obtains the barrier height $V_T=(k_A+k_c)(q_T^0)^2/2=46.35$ pNnm;
the force-free equilibrium positions are $q_A^0=0$, $q_T^0=0.3$ nm and  
$q_B^0=q_T^0+\sqrt{[2(V_T-V_B)-k_c(q_T^0)^2]/k_B}=1.0$ nm;\\
{\bf b:} Kinetic rates $k_{\rm on/off}(f)$ vs. the force $f$ in various approximations. 
The inset shows the ratio of the exact off-rate to the product approximation (\ref{tau.xy.app}) and the Kramers approximation, indicating the quality of the former.
\item[Fig.2 ] (color online)\\
{\bf a:}
Two examples of FE-curves as obtained from Brownian dynamics simulations.
The potential is of the shape shown in Fig.1a and the explicit parameters are the same as in Fig.1a.\\
{\bf b:}
Rupture force distributions for the pull mode, $p(f)$, and the relax mode, $p_{\rm rel}(f)$, as obtained from Brownian dynamics simulations (histograms) and from the solution of the master equation, eq.(\ref{ME.f}).\\
{\bf c:}
Dynamic strength from Brownian dynamics simulations (full lines) and from the Gaussian approximation, eq.(\ref{FvsX.res}) (dashed lines).
Upper panel: $\mu_c=300$ pN/s, lower panel: $\mu_c=3000$ pN/s.
\item[Fig.3 ] (color online)\\
{\bf a:}
Rupture force distributions for the pull mode ($p(f)$, upper panel) and the relax mode ($p_{\rm rel}(f)$, lower panel) versus force $f=\mu_c\cdot t$ for 
$\mu_c=10^2$, $10^3$, $10^4$, $10^5$ pN/s (for $p(f)$ from left to right and for 
$p_{\rm rel}(f)$ from right to left). 
The remaining parameters are the same as in Fig.1a.
Also shown are the distributions for the case of vanishing rebinding, i.e. $\koff^0=0$ for the pull mode and $\kon^0=0$ for the relax mode as the dash-dotted lines. The equilibrium rupture force distribution 
$p^{\rm eq}(f)\equiv p^{\rm eq}_{\rm rel}(f)$ is shown as the dotted line.\\
{\bf b:} Rupture and rejoining forces (upper panel) and variances (lower panel) as a function of the loading rate.
The full lines are the mean rupture force according to eq.(\ref{moments.gen}), $\lg f_r\rg$, and the dash-dotted lines represent the values obtained from the maximum of the rupture and rejoining force distribution, $f_{r,m}$. The dotted lines are the mean rupture forces without rebinding, 
$\lg f_r\rg_0$.
In the lower panel $\s_r=\sqrt{\lg f_r^2\rg-\lg f_r\rg^2}$ is shown including the case of vanishing rebinding (dotted lines).\\
{\bf c:}
Averaged FE-curves calculated in the Gaussian approximation, eq.(\ref{FvsX.res}), for pull mode (full lines) and relax mode (dash-dotted lines), $\mu_c=10^2$, $10^3$, $10^4$ pN/s. 
The maxima increase for the pull mode and decrease for relax mode. 
Also shown as the dotted line is the equilibrium curve, $\mu_c=0$. The two straight lines are the factors $(f-k_cq_X(f))$, $X=A$, $B$.
The inset shows the comparison to the dynamic strength $\hat F(f)$, cf. eq.(\ref{DS.def}), for the pull mode (dashed lines).\\
{\bf d:} Rupture forces $f_{r,m}$ (full lines), $F_m$ (dashed lines) and $\hat F_m$ (dotted lines) as a function of the loading rate. 
The inset shows the data in the pull mode with $F_m$ and $\hat F_m$ shifted in order to match $f_{r,m}$ at small $\mu_c$.
\item[Fig.4 ] (color online)\\
{\bf a:}
Rupture force $\lg f_r\rg$ versus loading rate in the local harmonic approximation (Kramers rates, 
eq.(\ref{koff.Kramers}), full lines), the Bell model (eq.(\ref{koff.Bell}), dashed lines) and for the cusp potential discussed in subsection A. 
The parameters for the Kramers rates in the local harmonic approximation are:
the local curvatures are $k_A=1300.0$ pN/nm, $k_B=48.49$ pN/nm and $k_T=10^4$ pN/nm;
the diffusion constant is $D=1442.75$ nm$^2$s; 
the potential heights are $V_A=0$, $V_B=33.12$ pNnm and $V_T=46.35$ pNnm;
the force-free equilibrium positions are $q_A^0=0$, $q_T^0=0.3$ nm and $q_B^0=1.0$ nm;
note that $V_A$, $V_B$, $V_T$, $q_A^0$, $q_T^0=$ and $q_B^0$ are the same as in Fig.1a;
the cantilever force constant was set to $k_c=30$ pN/nm.\\
{\bf b:}
Effect of variation of the parameters characterizing the transition state on $\lg f_r\rg$:
upper panel: Variation of the position $q_T^0$ for force constant $k_T=10^3$ pN/nm.
Also included as the dotted lines are the results for Bell rates in the pull mode for large 
$\mu_c$ for $q_T^0=0.3$, $0.4$ and $0.5$ nm.\\
lower panel: Variation of the curvature $k_T$ for constant position $q_T^0=0.4$ nm.
The remaining parameters are the same as in Fig.4a.\\
{\bf c:}
Mean rupture force for varying cantilever force constant $k_c$.
Upper panel: $\lg f_r\rg$ versus $\mu_c$ for different $k_c$.
Lower panel: $\lg f_r\rg^{\rm eq}$ versus $k_c$.
For comparison, also the results for the Bell model, eq.(\ref{fr.eq.Bell}), are shown as the dotted lines:
$Q_{BA}=\b k_B/(k_B+k_c)q_B^0$ (upper line), $Q_{BA}=\b q_B^0$ (lower line).
All other parameters are the same as in Fig.4a.
\item[Fig.5 ] (color online)\\
{\bf a:}
Rupture force distributions in the pull mode and the relax mode in the presence of a soft WLC linker for different loading rates $\mu_c$. The persistence length is fixed to be $l_p=0.4$ nm and the cantilever force constant used is $k_c=5$ pN/nm. The other parameters are the same as in Fig.4a. The different lines correspond to different contour lengths $L_c$ as indicated in the uppermost panel.
Here, $L_c=0$ means that no soft linker is included, cf. eq.(\ref{mu.WLC}) and the expression for $C_{WLC}$.\\
{\bf b:} 
$\lg f_r\rg^{\rm eq}$ versus $\mu_c$ for contour lengths $L_c=0$ (full lines) and $L_c=50$ nm (dashed lines). The other parameters are the same as in (a). 
Also included as the dotted line is the result from a calculation with an effective cantilever force constant $k_{\rm eff}=(k_c^{-1}+k_{WLC}^{-1})^{-1}$\cite{Hummer:2003p934} where 
$k_{WLC}=3/(2\b l_pL_c)\simeq 0.3105$ pN/nm is the force constant of the WLC-linker.
%
%greg
\end{description}
\newpage
\begin{figure}
\includegraphics[width=17cm]{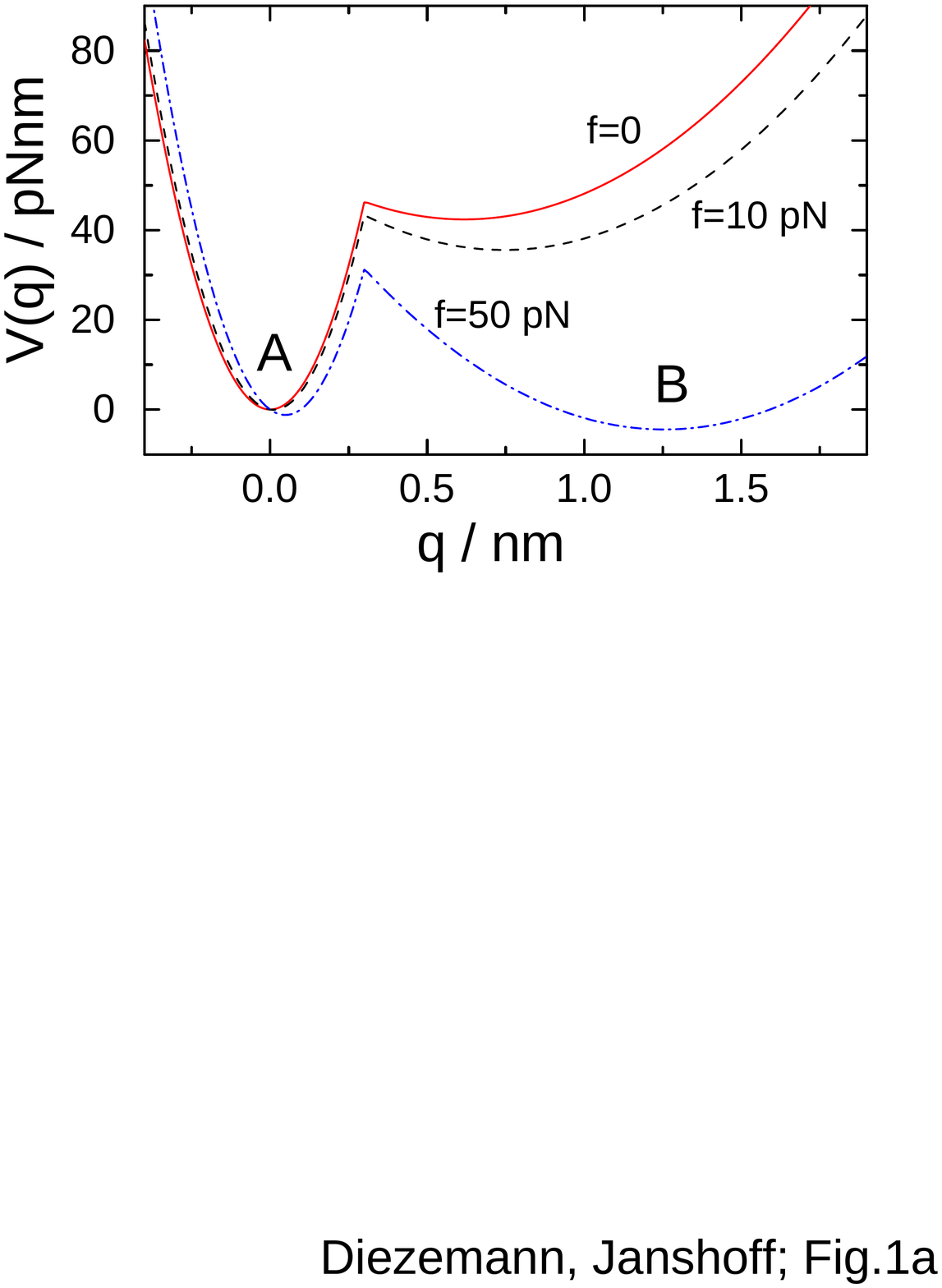}
\end{figure}
\newpage
\begin{figure}
\includegraphics[width=17cm]{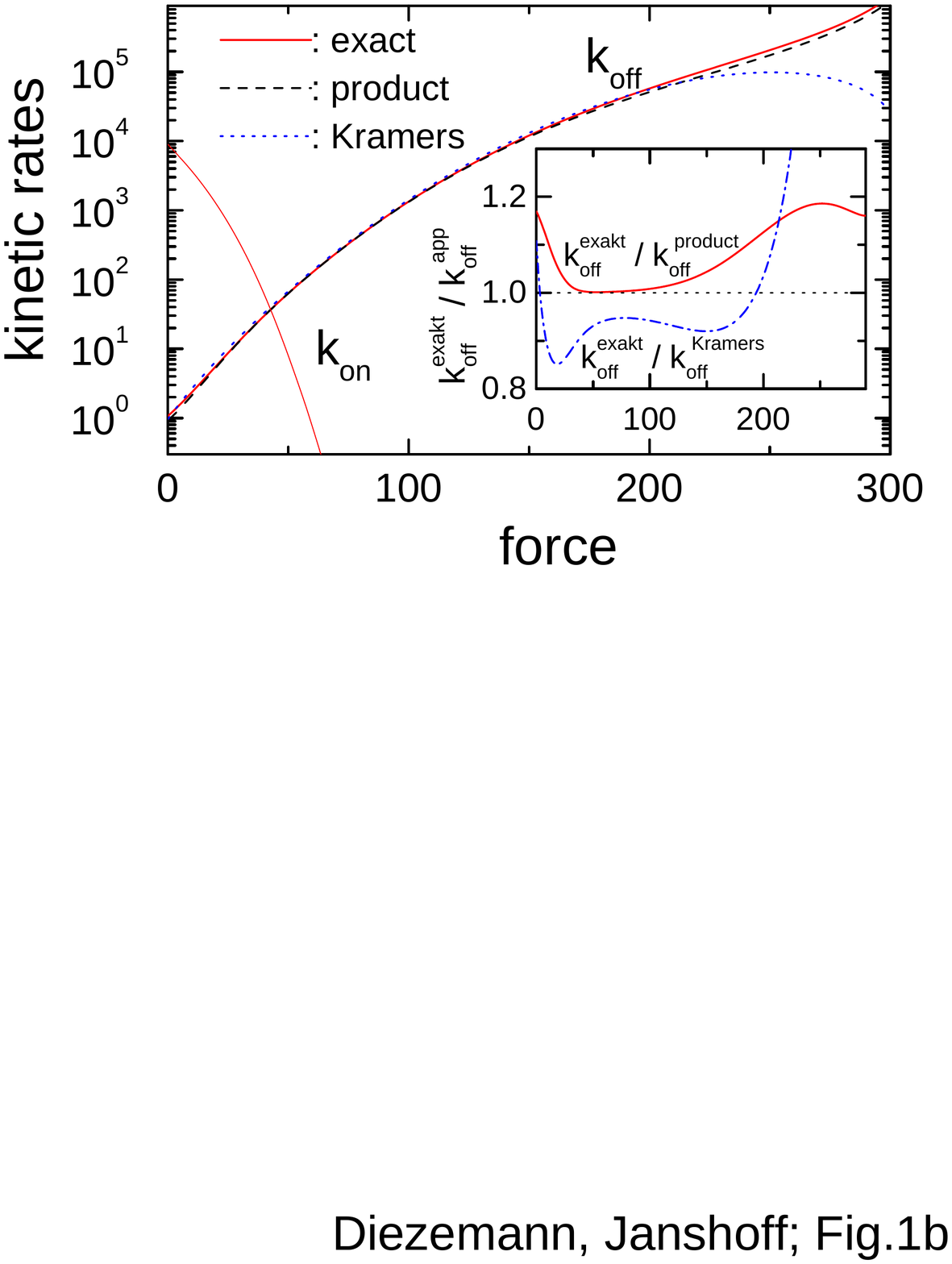}
\end{figure}
\newpage
\begin{figure}
\includegraphics[width=17cm]{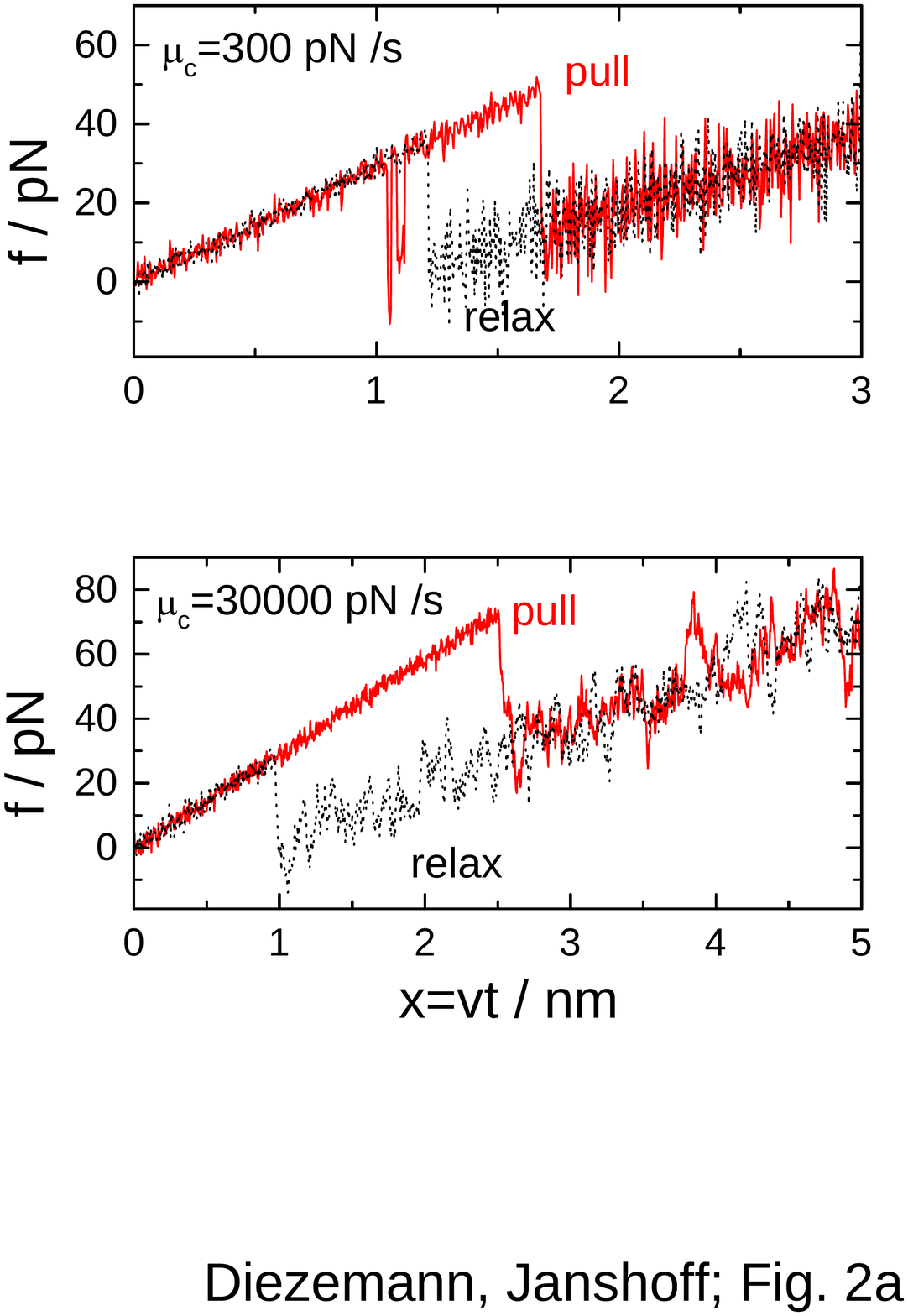}
\end{figure}
\newpage
\begin{figure}
\includegraphics[width=17cm]{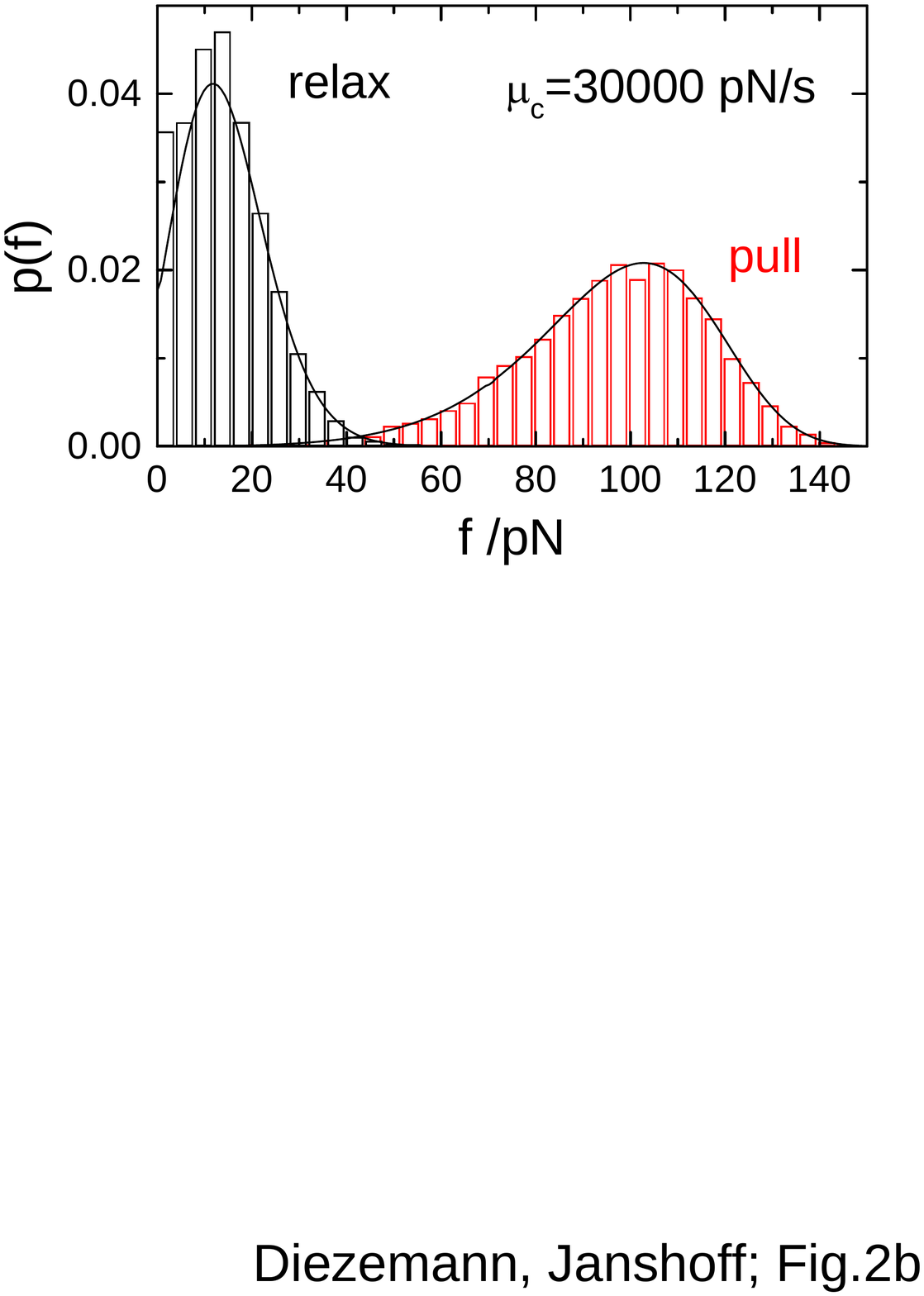}
\end{figure}
\newpage
\begin{figure}
\includegraphics[width=17cm]{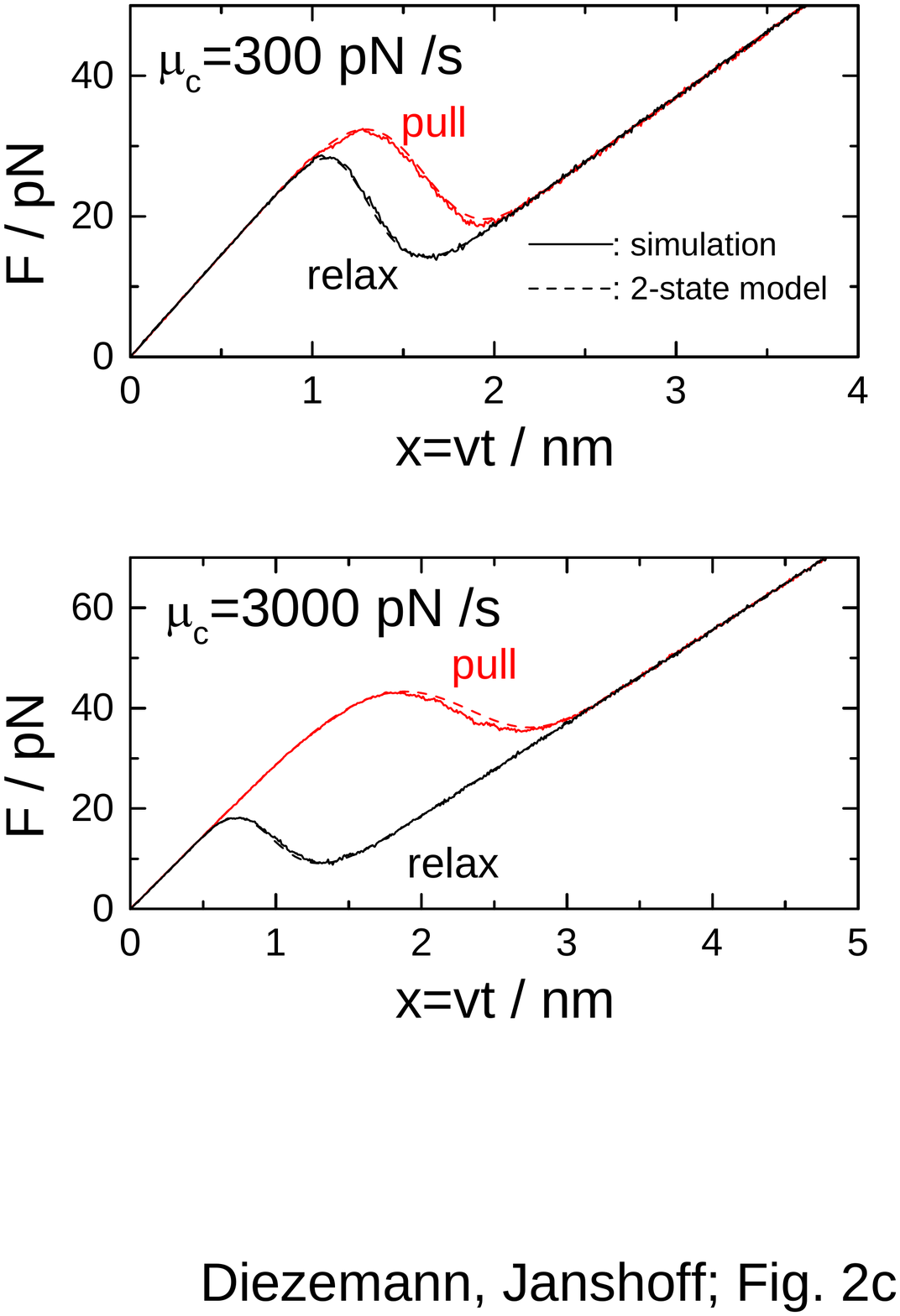}
\end{figure}
\newpage
\begin{figure}
\includegraphics[width=17cm]{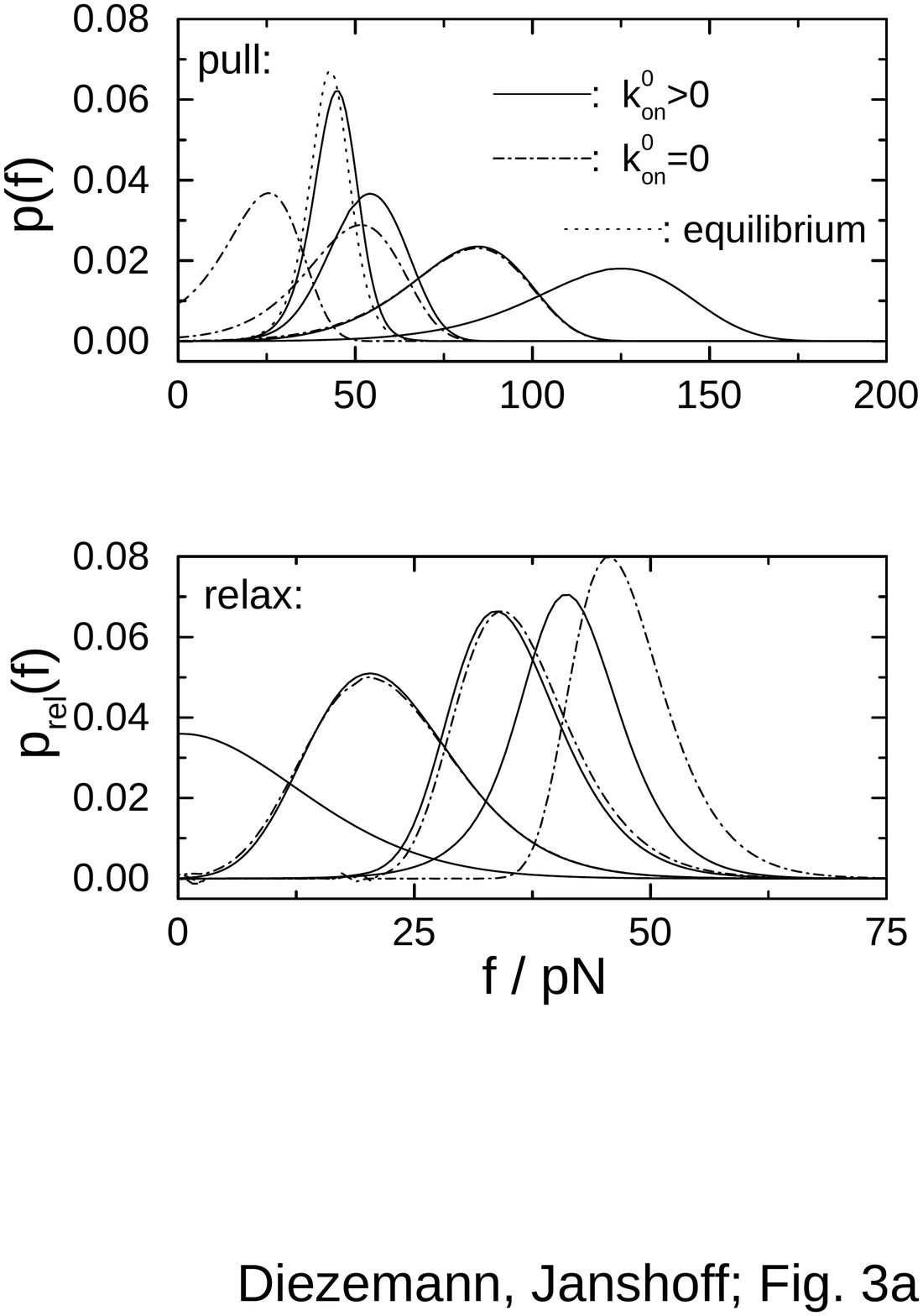}
\end{figure}
\newpage
\begin{figure}
\includegraphics[width=17cm]{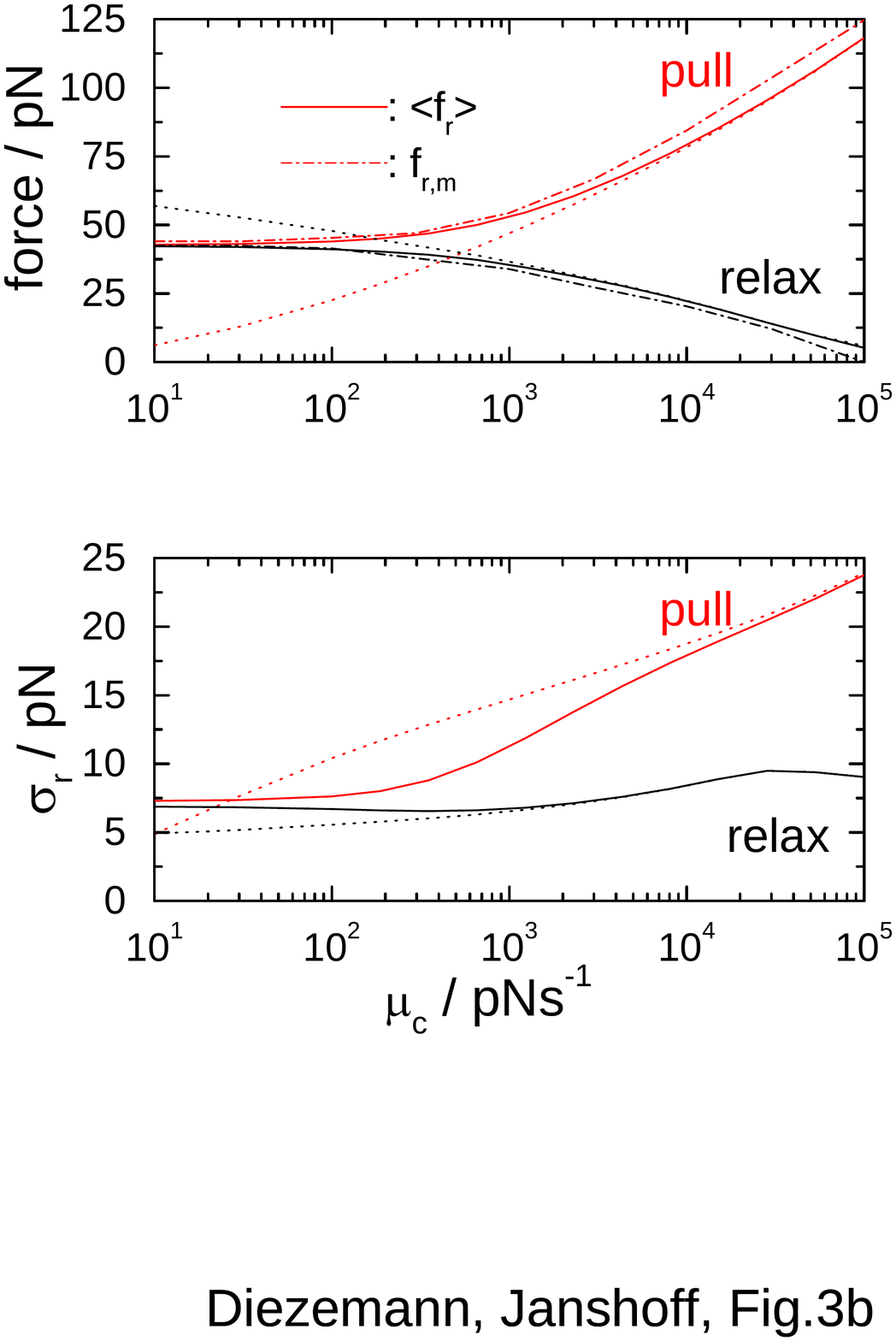}
\end{figure}
\newpage
\begin{figure}
\includegraphics[width=17cm]{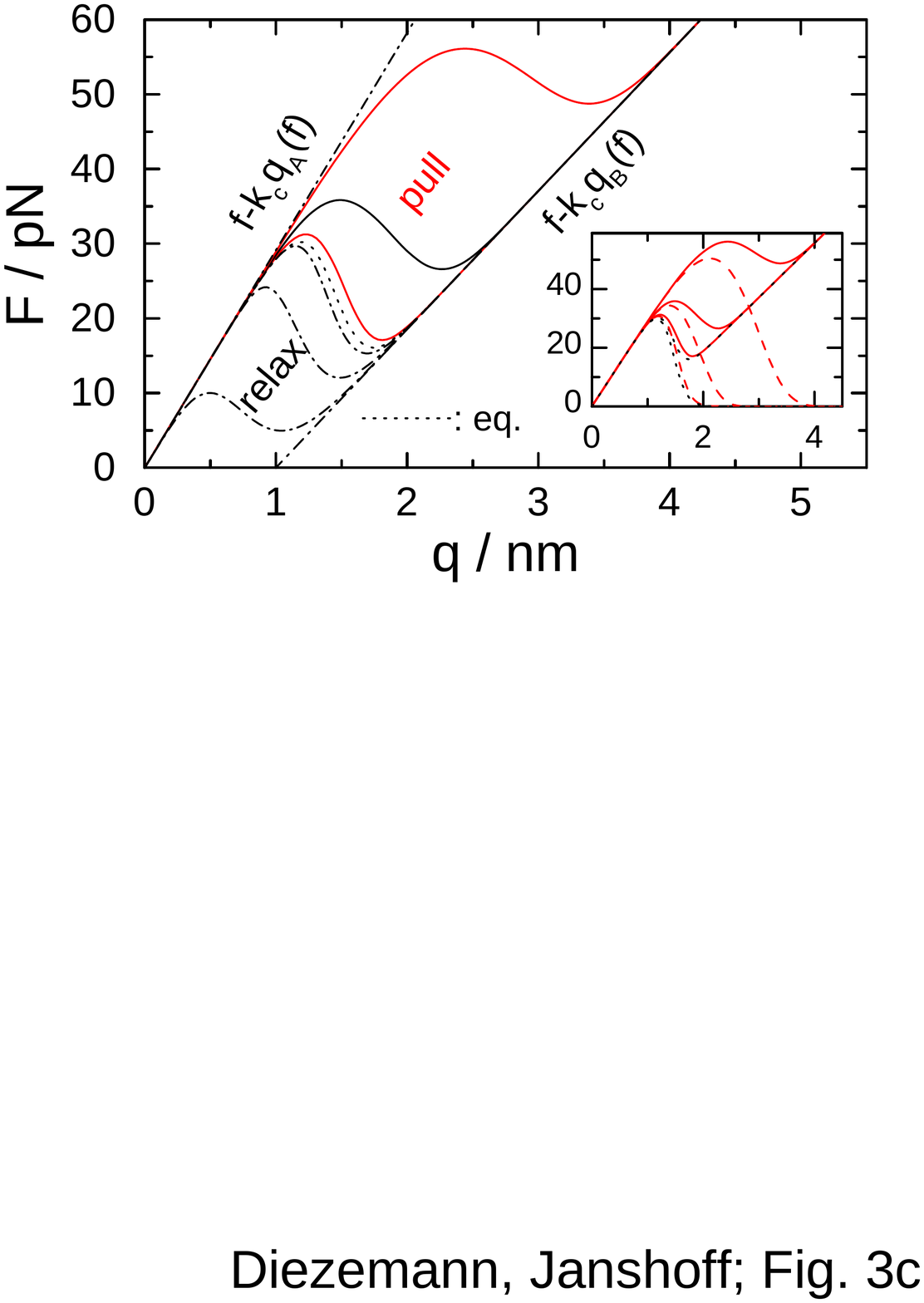}
\end{figure}
\newpage
\begin{figure}
\includegraphics[width=17cm]{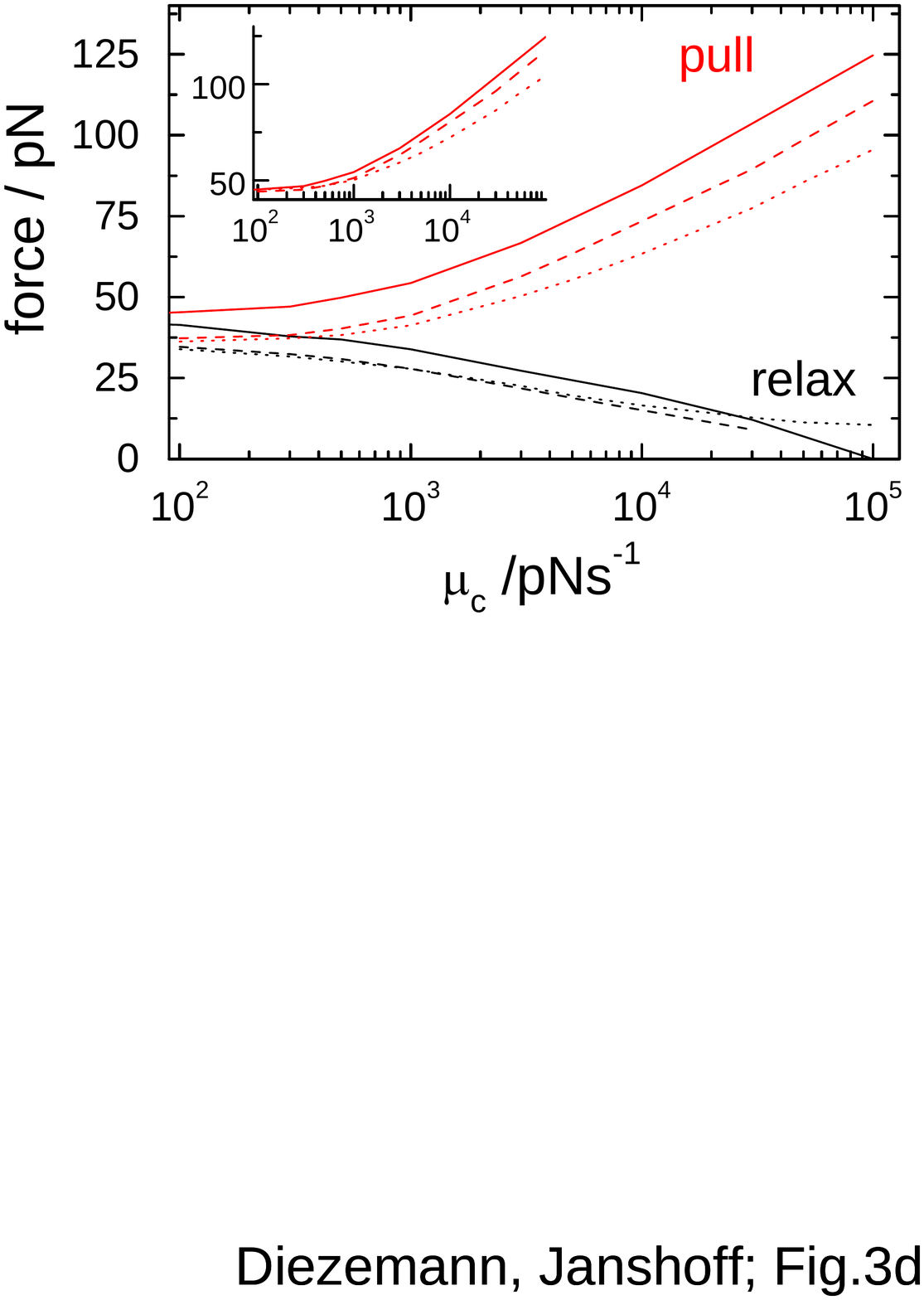}
\end{figure}
\newpage
\begin{figure}
\includegraphics[width=17cm]{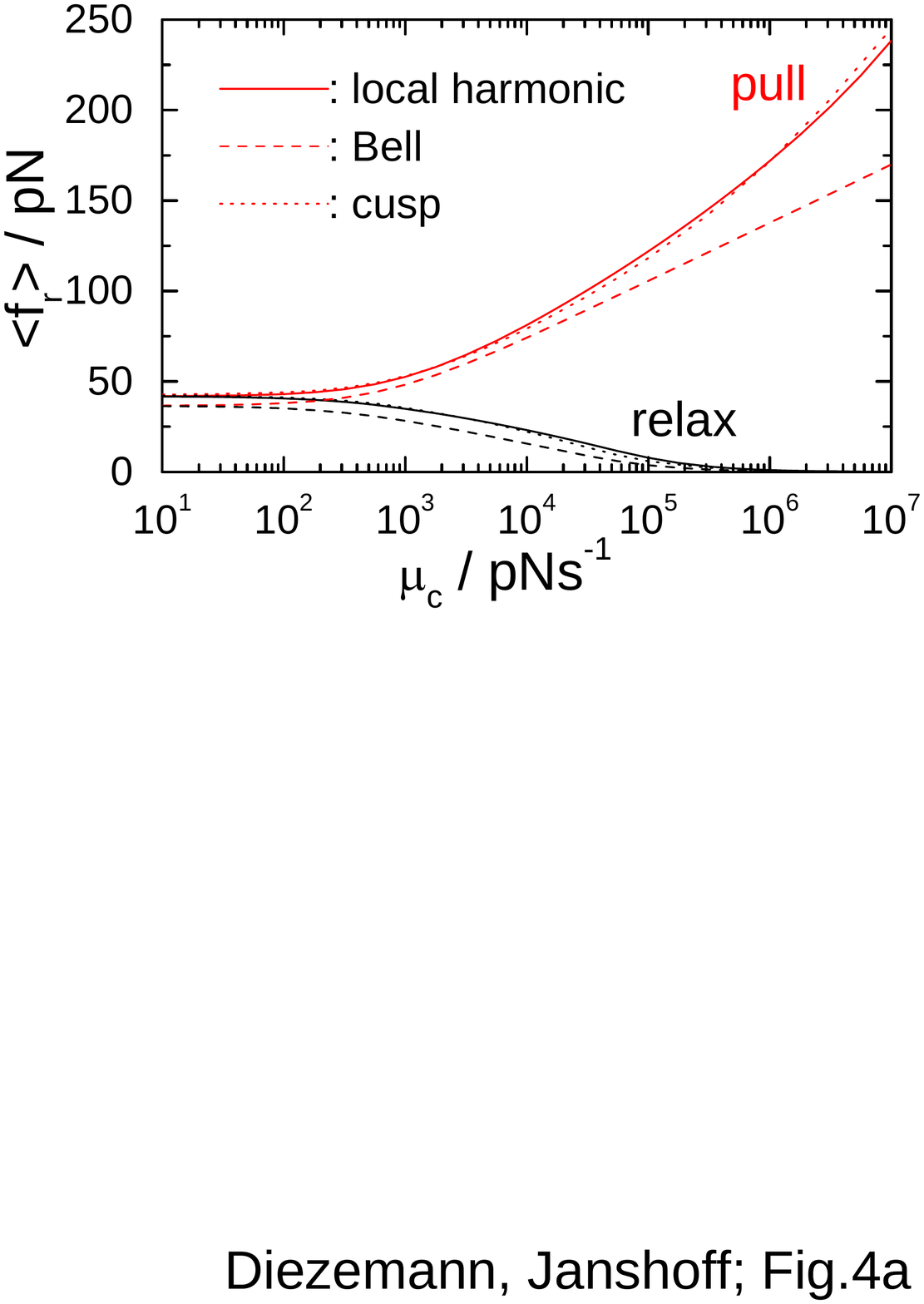}
\end{figure}
\newpage
\begin{figure}
\includegraphics[width=17cm]{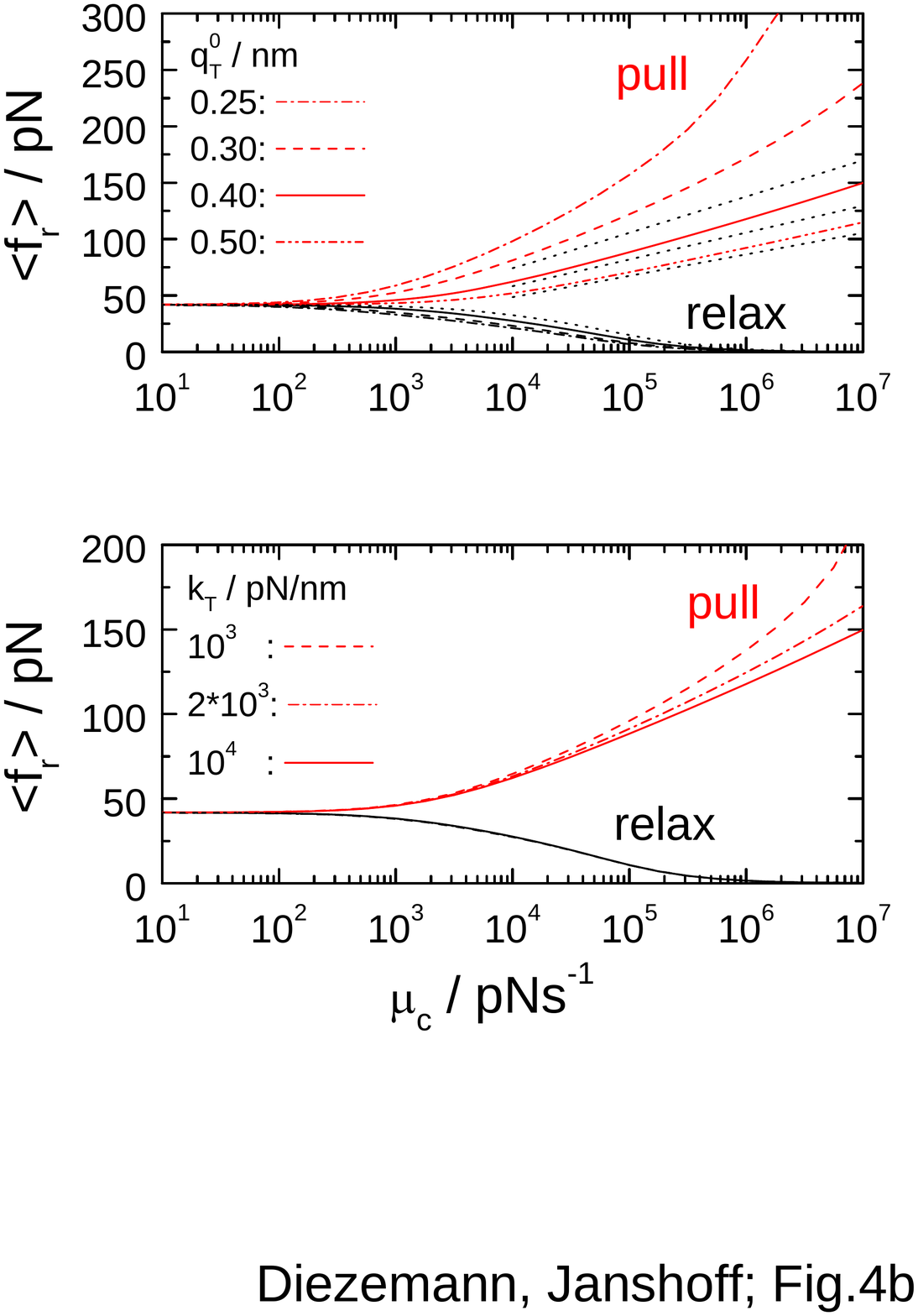}
\end{figure}
\newpage
\begin{figure}
\includegraphics[width=17cm]{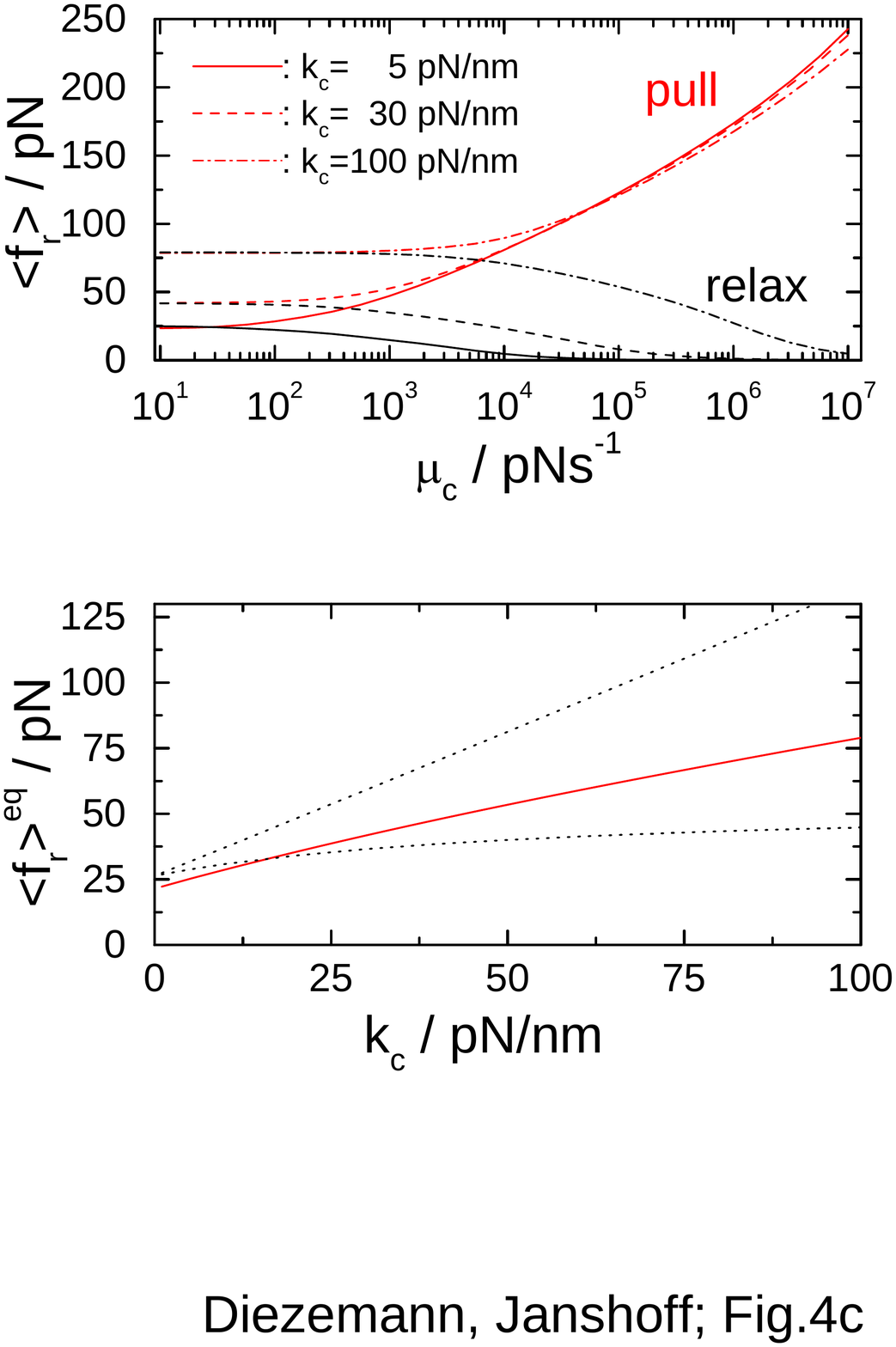}
\end{figure}
\newpage
\begin{figure}
\includegraphics[width=17cm]{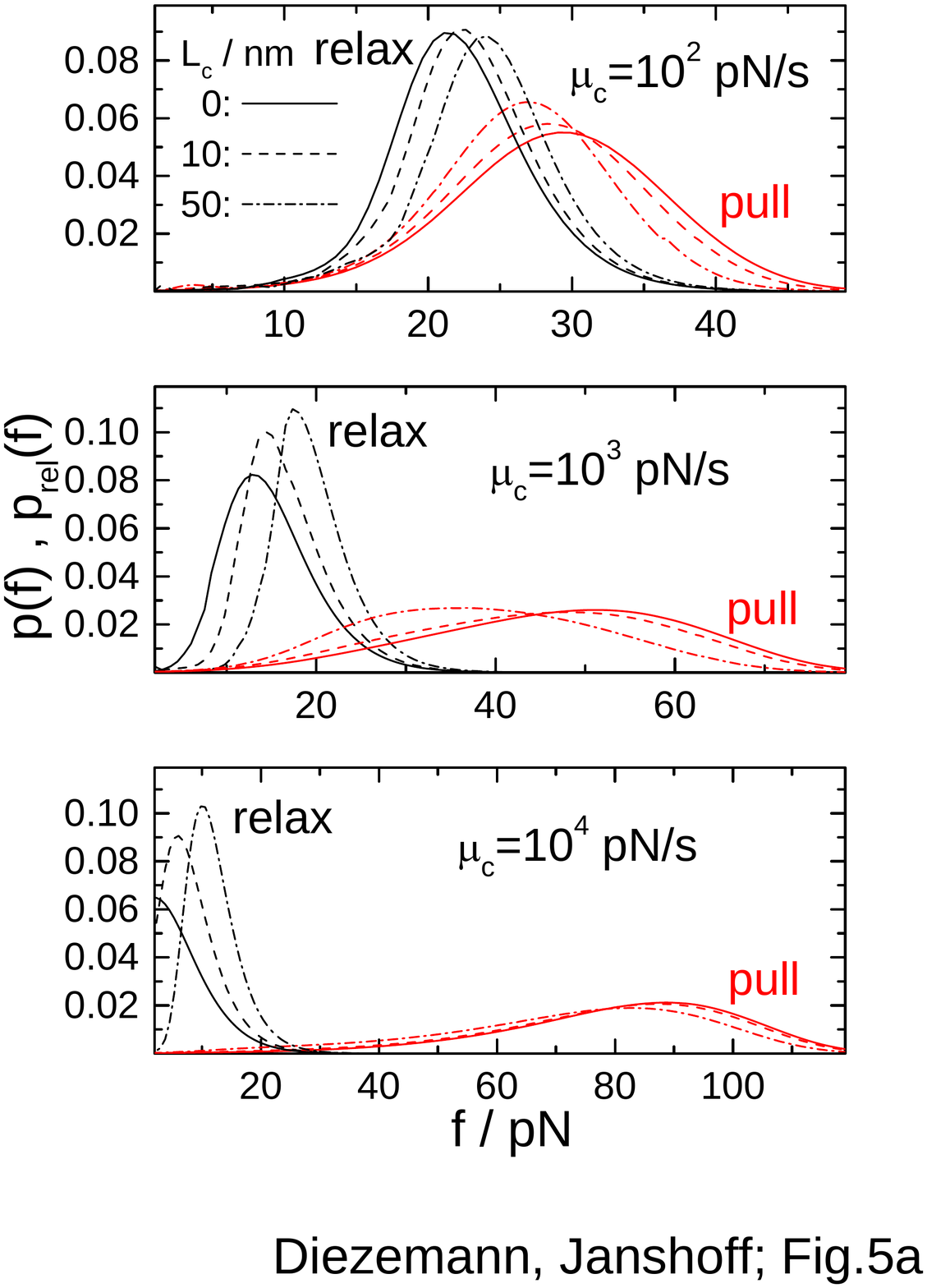}
\end{figure}
\newpage
\begin{figure}
\includegraphics[width=17cm]{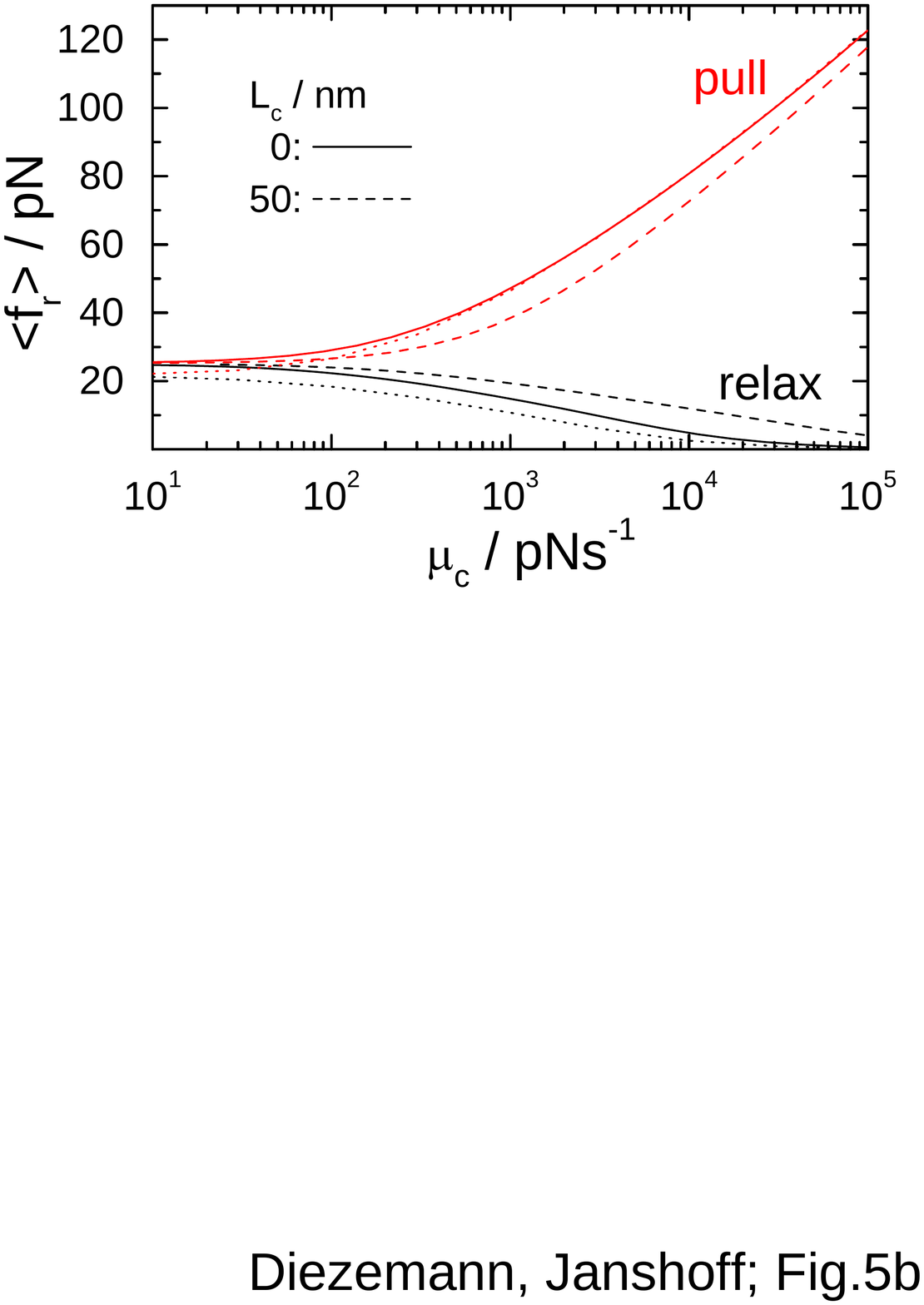}
\end{figure}
%

%\newpage
\end{document}